\documentclass[12pt]{JHEP3}
\usepackage{amsmath}
\usepackage{amssymb}
\usepackage{epsfig}
\usepackage{longtable}

\ifpdf
\else
\usepackage{colordvi}
\fi

\numberwithin{equation}{section}

\title{Fano hypersurfaces and Calabi-Yau 
supermanifolds}

\author{
Richard S.~Garavuso, Maximilian Kreuzer 
\\
Institut f\"{u}r Theoretische Physik, Technische
Universit\"{a}t Wien
\\
Wiedner Hauptstrasse 8-10/136, A-1040 Vienna, Austria 
\\
E-mail: \email{garavuso@hep.itp.tuwien.ac.at}, 
\email{kreuzer@hep.itp.tuwien.ac.at}  
}

\author{
Alexander Noll
\\
Institut f\"{u}r Theoretische Physik, Technische Universit\"{a}t Graz
\\
Petersgasse 16/II, A-8010 Graz, Austria
\\
E-mail: \email{hava@sbox.tugraz.at}
}

\abstract{In this paper, we study the geometrical interpretations 
associated with Sethi's proposed general correspondence between 
$ \mathcal{N} = 2 $ Landau-Ginzburg orbifolds with integral $ \hat{c} $ 
and $ \mathcal{N} = 2 $ nonlinear sigma models.
We focus on the supervarieties associated with $ \hat{c} = 3 $ Gepner 
models.
In the process, we test a conjecture regarding the superdimension of the 
singular locus of these supervarieties.
The supervarieties are defined by a hypersurface 
$ \widetilde{W} = 0 $ in a weighted superprojective space and have 
vanishing super-first Chern class.
Here, $ \widetilde{W} $ is the modified superpotential obtained by 
adding as necessary to the Gepner superpotential a boson mass term 
and/or fermion bilinears so that the superdimension of the supervariety 
is equal to $ \hat{c} $. 
When Sethi's proposal calls for adding fermion bilinears, setting the 
bosonic part of $ \widetilde{W} $ (denoted by $ \widetilde{W}_{bos} $) 
equal to zero defines a Fano hypersurface embedded in a weighted 
projective space.
In this case, if the Newton polytope of $ \widetilde{W}_{bos} $ admits a 
nef partition, then the Landau-Ginzburg orbifold
can be given a geometrical interpretation as a nonlinear sigma model on
a complete intersection Calabi-Yau manifold.
The complete intersection Calabi-Yau manifold should be equivalent 
to the Calabi-Yau supermanifold prescribed by Sethi's proposal.
}


\preprint{}

\begin{document}

\section{\label{Introduction}Introduction}

Mirror symmetry \cite{Mirror} is a duality between string theories 
propagating on distinct but mirror target spaces.  
Consider a string theory compactified on a Calabi-Yau manifold $ X $ 
which is related by mirror symmetry to a string theory compactified on a 
Calabi-Yau manifold $ Y $.
The mirror map relates the Hodge numbers of $ X $ and $ Y $  by
$ h^{p,q}_X = h^{\mathcal{D}-p,q}_Y $,
where $ \mathcal{D} $ is the complex dimension of $ X $ and $ Y $.
Thus, the mirror map identifies the complex structure moduli space of
$ X $ with the K\"{a}hler moduli space of $ Y $ and vice-versa.

A \emph{rigid} Calabi-Yau manifold has no complex structure moduli.  
The mirror of such a manifold has no K\"{a}hler moduli and hence cannot 
be a K\"{a}hler manifold in the conventional sense.  
Thus, Calabi-Yau manifolds cannot be the most general geometrical 
framework for understanding mirror symmetry.
The first progress towards generalizing this framework came when 
Schimmrigk \cite{Schimmrigk} suggested that higher-dimensional Fano 
varieties could provide mirrors for rigid Calabi-Yau manifolds.
The name ``generalized Calabi-Yau'' was introduced by Candelas et al.
\cite{CandelasDerrickParkes:Generalized} for these mirror manifolds.
Later progress came when Sethi \cite{Sethi} proposed a general 
correspondence between $ \mathcal{N} = 2 $ Landau-Ginzburg orbifolds 
with integral $ \hat{c} \equiv c/3 $ (where $ c $ is the central charge) 
and $ \mathcal{N} = 2 $ nonlinear sigma models.
Here, the target space of the nonlinear sigma model is either a 
Calabi-Yau manifold or a Calabi-Yau \emph{supermanifold}.
Using this proposal, Sethi argued that the mirror of a rigid Calabi-Yau 
manifold is a Calabi-Yau supermanifold and hence mirror symmetry should 
be viewed as a relation among Calabi-Yau manifolds and Calabi-Yau 
supermanifolds alike.
The bodies of the supermanifolds are the Fano varieties mentioned above.
Witten \cite{Witten:Phases} described $ \mathcal{N} = 2 $ nonlinear
sigma models on Calabi-Yau manifolds and  $ \mathcal{N} = 2 $
Landau-Ginzburg orbifolds as being different phases of 
$ \mathcal{N} = 2 $ gauged linear sigma models.
Sethi's work and \cite{Schwarz:Sigma} have led others \cite{others} to 
study $ \mathcal{N} = 2 $ gauged linear sigma models which have a phase 
described by an $ \mathcal{N} = 2 $ nonlinear sigma model on a 
Calabi-Yau supermanifold. 
The gauged linear sigma model framework allows the argument establishing 
a correspondence between the nonlinear sigma model and the 
Landau-Ginzburg orbifold to be made more robust.

A Calabi-Yau supermanifold $ M $ obtained from Sethi's proposed 
correspondence would be realized by resolving the singularities of a 
supervariety $ \mathcal{M} $ embedded in a weighted superprojective 
space 
\begin{equation}
{\bf WSP}^{(n|2m)} \equiv
  \mathbf{WSP}(n_{z_1},\ldots,n_{z_{n+1}}|n_{\eta_1},\ldots,n_{\eta_{2m}}) 
  \, .
\end{equation}
Here, $ \mathcal{M} $ is defined by the zero locus of a 
transverse\footnote{Transverse $ \widetilde{W} $ means that 
$ \widetilde{W} = 0 $ and $ d \widetilde{W} = 0 $ have no common 
solution except at the origin.}, 
quasihomogeneous superpotential 
$ \widetilde{W} = \widetilde{W}(z_{\mu}; \eta_{\alpha}) $, where 
\begin{equation}
\label{WeightedProjectiveIdentification}
z_{\mu} \simeq  \lambda^{  n_{ z_{\mu} }  } z_{\mu}  \ 
(\mu = 1,\ldots,n+1) \, , 
\quad 
\eta_{\alpha} \simeq \lambda^{  n_{ \eta_{\alpha} }  } \eta_{\alpha} \ 
(\alpha = 1,\ldots,2m)  
\end{equation}
are homogeneous bosonic and fermionic coordinates of weights
$ n_{z_{\mu}} $ and $ n_{ \eta_{\alpha} } $, respectively.
Since $ \widetilde{W} $ is quasihomogeneous, it satisfies
\begin{equation}
\widetilde{W}
  ( \lambda^{  n_{ z_{\mu} }  } z_{\mu}; 
    \lambda^{  n_{ \eta_{\alpha} }  } \eta_{\alpha} )
  = \lambda^d  \, \widetilde{W}( z_{\mu}; \eta_{\alpha} ) \, ,
\end{equation}
where $ d $ is the degree of
quasihomogeneity.
The superpotential $ \widetilde{W} $ is obtained from the transverse, 
quasihomogeneous superpotential $ W = W(\Phi_a) $ of an 
$ \mathcal{N} = 2 $ Landau-Ginzburg model with integral $ \hat{c} $ by 
truncating each chiral superfield $ \Phi_a \ (a=1,\ldots,N) $ to 
its lowest bosonic component $ \phi_a $, setting $ \phi_a = z_a $, and 
then adding boson mass terms 
$ z^2_{N+1} + \cdots + z^2_{n+1} $
and/or fermion bilinears 
$ \eta_1 \eta_2 + \cdots + \eta_{2m-1} \eta_{2m} $
to $ W $ so that
\begin{equation}
\label{superdimension}
\widetilde{\mathcal{D}} \equiv (n + 1) - 2m - 2 = \hat{c} \, .
\end{equation}
Here, $ \widetilde{\mathcal{D}} $ is the \emph{superdimension} 
of the Calabi-Yau supermanifold.
The condition (\ref{superdimension}) allows a change of variables with 
constant Jacobian to be made such that one of the new variables appears 
only linearly in the modified superpotential $ \widetilde{W} $.  
The change of variables is not one-to-one, so the modified theory must 
be orbifolded by the diagonal subgroup of its phase symmetries.
Integrating the linear new variable out of the path integral for the 
modified action as a Lagrange multiplier yields a super-delta function 
constraint which  corresponds to $ \mathcal{M} $ having vanishing 
super-first Chern class.

When $ m \neq 0 $, setting the bosonic part of $ \widetilde{W} $ 
(denoted by $ \widetilde{W}_{bos} $) equal to zero defines a \emph{Fano 
hypersurface} $ \mathcal{F} $  embedded in a weighted projective space
\begin{equation}
 {\bf WP}^n \equiv {\bf WP}( n_{z_1},\ldots,n_{z_{n+1}} ) \, .
\end{equation}
In this case, if the Newton polytope $ \Delta $ corresponding to 
$ \widetilde{W}_{bos} $ admits a \emph{nef partition} 
$  \Delta = \Delta_1 + \cdots + \Delta_r $, then the Landau-Ginzburg 
orbifold can be given a geometrical interpretation as a nonlinear sigma 
model on a complete intersection Calabi-Yau manifold defined by equations 
$ f_i = 0 \ (i=1,\ldots,r) $ \cite{nef}.
Here, $ \Delta_i $ is the Newton polytope of $ f_i $.
The complete intersection Calabi-Yau manifold should be equivalent (in 
the sense of \cite{Schwarz:Sigma}) to the Calabi-Yau supermanifold 
prescribed by Sethi's proposed correspondence.
When $ m=0 $, the constraint $ \widetilde{W} = 0 $ defines a Calabi-Yau 
variety $ \mathcal{X} $ embedded in $ {\bf WP}^n $. 
Resolving $ \mathcal{X} $ would yield a Calabi-Yau manifold $ X $ of 
complex dimension
\begin{equation}
\mathcal{D} \equiv (n + 1) - 2 = \hat{c} \, .
\end{equation}
The complex dimension of the singular locus of $ \mathcal{X} $ satisfies
\cite{dimsinglocus}
\begin{equation}
\label{BosonicSingularLocusConstraint}
0 \leq \mathrm{dim}(  \mathrm{Sing}( \mathcal{X} )  ) 
  \leq \mathcal{D} - 2 \, , 
\quad
\mathrm{Sing}(\mathcal{X})  
  = \mathcal{X} \cap \mathrm{Sing}( \mathbf{WP}^n ) \, .
\end{equation}
To obtain a Calabi-Yau supermanifold $ M $ by resolving the singularities of 
$ \mathcal{M} $, one might infer from the discussion in \cite{Sethi} 
that the superdimension of the singular locus of $ \mathcal{M} $ must satisfy
\begin{equation}
\label{SingularLocusConstraint}
\mathrm{sdim}( \mathrm{Sing}(\mathcal{M}) )
  \leq \widetilde{\mathcal{D}} - 1 \, ,
\quad
\mathrm{Sing}(\mathcal{M})
  = \mathcal{M} \cap \mathrm{Sing}( \mathbf{WSP}^{(n|2m)} ) \, .
\end{equation}
The result (\ref{BosonicSingularLocusConstraint}) and the 
equivalence discussed in \cite{Schwarz:Sigma} suggests the following 
stronger conjecture: 
\newtheorem{conjecture}{Conjecture}[section]
\begin{conjecture}
\label{ConjecturedSingularLocusConstraint}
To obtain a Calabi-Yau supermanifold $ M $ by resolving the 
singularities of $ \mathcal{M} $, the superdimension of the singular 
locus of $ \mathcal{M} $ must satisfy
\begin{equation*}
\mathrm{sdim}( \mathrm{Sing}(\mathcal{M}) ) 
  \leq \widetilde{\mathcal{D}} - 2 \, ,
\quad
\mathrm{Sing}(\mathcal{M})
  = \mathcal{M} \cap \mathrm{Sing}( \mathbf{WSP}^{(n|2m)} ) \, .
\end{equation*}
\end{conjecture}

In this paper, we will test the above conjecture for  
$ \widetilde{\mathcal{D}} = 3 $.
This will be achieved by studying the geometrical interpretations 
prescribed by Sethi's proposed correspondence for \emph{Gepner models} 
\cite{originalGepner} with $ \hat{c} = 3 $.
Since $ \widetilde{W} $ is quasihomogeneous of degree $ d $, the weights 
of the fermions in each fermion bilinear 
$ \eta_{2k-1} \eta_{2k} \ (k=1,\ldots,m) $ must satisfy 
\begin{equation}
\label{QuasihomogeneityConstraint}
n_{\eta_{2k-1}} + n_{\eta_{2k}} = d \, .
\end{equation}
Requiring either the singular locus constraint 
(\ref{SingularLocusConstraint}) or Conjecture 
\ref{ConjecturedSingularLocusConstraint} to hold further restricts the 
fermionic weights.
We have written a computer program which allows these restrictions to be 
implemented.
In principle, one could determine the fermionic weights by requiring 
agreement between the Hodge diamond of the Landau-Ginzburg orbifold and 
the Hodge diamond of $ \mathcal{M} $.
The former Hodge diamond can be computed using the techniques of 
\cite{LGOHodge} whereas insight into the structure of the latter Hodge 
diamond can be obtained from the heuristic approach of \cite{Sethi} 
based on orbifold considerations \cite{Zaslow:Topological}.
We will compare the fermionic weights obtained in this way with those 
obtained from our computer program.

This paper is organized as follows:  
In Section \ref{Gepner/NLSM}, we discuss the application of Sethi's 
proposed Landau-Ginzburg orbifold/nonlinear sigma model correspondence 
to Gepner models.
In Section \ref{Fermionic}, we describe how the singular locus 
constraint (\ref{SingularLocusConstraint}) restricts the fermionic 
weights and work through an example.
It is a trivial step to replace (\ref{SingularLocusConstraint}) with 
Conjecture \ref{ConjecturedSingularLocusConstraint} in our computer 
program.
The analysis of Section \ref{Analysis} compares the fermionic weights 
obtained from our computer program with those obtained from the 
cohomological approach described in the previous paragraph.
Several examples are included which highlight the similarities and 
differences.
Concluding remarks are given in Section \ref{Conclusion}.
Finally, in the Appendix, we tabulate the families of hypersurfaces 
associated with our supervarieties. 
Here, the fermionic weights are determined with our computer program by 
requiring  (\ref{SingularLocusConstraint}) and
(\ref{QuasihomogeneityConstraint}) to be satisfied.
For each hypersurface family, the Hodge numbers
$ h^{1,1} $ and $ h^{2,1} $ and the Euler number of the associated
Landau-Ginzburg orbifold are given.
When the Newton polytope corresponding to $ \widetilde{W}_{bos} $ admits 
a nef partition, this is indicated.
We also indicate when the Newton polytope of $ \widetilde{W}_{bos} $ is 
nonreflexive Gorenstein.

\section{\label{Gepner/NLSM}Gepner/NLSM correspondence}

The worldsheet action for an $ \mathcal{N} = 2 $ Landau-Ginzburg model 
is \cite{VafaWarner:Catastrophes}
\begin{equation}
\label{LGaction}
S = \int d^2 {\mathtt z} \, d^4 \theta \,
    K\left(\Phi_a, \overline{\Phi}_a \right)
    + \left( 
            \int d^2 {\mathtt z} \, d^2 \theta \, W(\Phi_a) + c.c. 
       \right),
\end{equation}
Here, the integral involving the K\"{a}hler potential $ K $ is called 
the $ D $-term, the integral involving the superpotential $ W $ is 
called 
the $ F $-term, and $ \Phi_a \ (a=1,\ldots,N) $ are chiral superfields.
The $ D $-term contains only irrelevant operators whereas the $ F $-term
contains relevant operators.
Thus, the superpotential defines a universality class under
renormalization group flow.
Requiring the superpotential to be transverse and quasihomogeneous 
is believed to ensure the existence of a unique, nontrivial IR fixed 
point which is conformally invariant.
At this fixed point, the action (\ref{LGaction}) provides a Lagrangian 
description of an $ \mathcal{N} = 2 $ minimal model 
\cite{VafaWarner:Catastrophes,Martinec:Algebraic,CecottiGiradelloPasquinucci} 
with
\begin{equation}
\hat{c} = 2 \sum_{a=1}^{N} ( \frac{1}{2} - q_{\Phi_a} ) \, , 
\quad
q_{\Phi_a} \equiv n_{\Phi_a} / d \, .
\end{equation} 

The 10,839 transverse, quasihomogeneous Landau-Ginzburg superpotentials 
corresponding to  $ \mathcal{N}=2 $ superconformal theories with 
$ \hat{c} = 3 $ were classified in \cite{KreuzerSkarke:NoMirror}.
A subset of these correspond to Gepner models with $ \hat{c} = 3 $.
A Gepner model \cite{originalGepner} is a string model constructed 
as an orbifold of a tensor product of $ \mathcal{N}=2 $ minimal models.
The central charge of the $ i $th $ \mathcal{N}=2 $ minimal model in the
tensor product is
\begin{equation}
c_i = \frac{3k_i}{k_i + 2}
\quad (k_i = 1,2,\ldots) \, ,
\end{equation}
where $ k_i $ is the \emph{level} of the $ \mathcal{N} = 2 $
superconformal algebra \cite{N=2SuperconformalAlgebra}.
To obtain an anomaly-free compactification to $ D $ spacetime dimensions
$ (D < 10 $, even), the internal contribution to the central charge must
be
\begin{equation}
c = \sum_i c_i = \frac{3}{2}(10-D) \, .
\end{equation}

The work of \cite{GreeneVafaWarner:Calabi-Yau,FuchsKlemmScheichSchmidt}
associated Calabi-Yau manifolds to a large class of Gepner models.
Sethi's proposal \cite{Sethi} prescribes a geometrical interpretation 
for \emph{all} Gepner models.  
This prescription is as follows:
\begin{enumerate}

\item{Start by associating superpotential terms 
\cite{Martinec:Algebraic}
\begin{itemize}

\item{ $ W_i= x^{k_i + 2}_i $ to $ A $-models of level $ k_i $,}

\item{ $ W_i = x^{\frac{k_i}{2} + 1}_i + x_i y^2_i $  to
$ D $-models of level $ k_i $ (even),}

\item{ $ W_i= x^3_i + y^4_i $,  $ W_i = x^3_i + x_i y^3_i $, and
$ W_i = x^3_i + y^5_i $ to $ E_6 $-, $ E_7 $-, and $ E_8 $-models of
level $ k_i = 10,16,28 $, respectively.}

\end{itemize}
The tensor product of $ r $ subtheories yields a transverse, 
quasihomogeneous  Gepner superpotential
\begin{equation}
\label{GepnerSuperpotential}
W = \sum_{i=1}^r W_i
  = \sum_{i=1}^r
    \left(
          x_i^{l_{x_i}} + x^{\overline{l}_{x_i}}_i y^{l_{y_i}}_i
    \right).
\end{equation}
For $ A $-models, $ y_i = 0 $ and $ l_{x_i} = k_i + 2 $.
For $ D $-models, $ l_{x_i} = k_i / 2  + 1 $,
$ \overline{l}_{x_i} = 1 $, and $ l_{y_i} = 2 $.
For $ E_6 $-, $ E_7 $-, and $ E_8 $-models,
$ l_{x_i} = 3,3,3 $,
$ \overline{l}_{x_i} = 0, 1, 0 $, and
$ l_{y_i} = 4,3,5 $, respectively.
In all cases,
\begin{equation}
\label{l-relation1}
n_{x_i} l_{x_i} = d \, .
\end{equation}
Additionally, for $ D $- and $ E $-models,
\begin{equation}
\label{l-relation2}
n_{x_i} \overline{l}_{x_i} + n_{y_i} l_{y_i} = d \, .
\end{equation}
The $ x_i $ and nonzero $ y_i \ (i=1,\ldots,r) $ are identified with 
the $ z_a \ (a=1,\ldots,N) $ described in the Introduction according to 
the convention $ z_1 = x_1 $,
$$ z_2 = \left\{
            \begin{array}{l}
             y_1 \quad (y_1 \neq 0) \\
             x_2 \quad (y_1 = 0) \, ,          
            \end{array}  \right. $$
and so on.}         

\item{Add as necessary to $ W $ a single\footnote{There are three 
possibilities for an arbitrary Landau-Ginzburg model with integral $ 
\hat{c} $:
\begin{itemize}

\item{For $ \hat{c} = N -2 $, no fields need to be
added.}

\item{For $ \hat{c} > N - 2 $, boson mass terms
$ z^{2}_{N + 1} + \cdots + z^2_{n+1} $ are required.}

\item{For $ \hat{c} < N - 2 $, the condition that $ \hat{c} $
be integral implies that the sum of the charges
$ \sum_{a=1}^{N} q_{z_a} $ can be either integral or half-integral.
The former requires adding fermion bilinears
$ \eta_1 \eta_2 + \cdots + \eta_{2m-1} \eta_{2m} $
whereas the latter requires adding a single boson mass term
$ z^2_{n+1} $ and fermion bilinears
$ \eta_1 \eta_2 + \cdots + \eta_{2m-1} \eta_{2m} $.}  

\end{itemize}
For Gepner models, $ \hat{c} $ is never
greater than $ N - 2 $ by more than 1.  Thus, it is never necessary to 
add more than one boson mass term to the Gepner superpotential.
}  
boson mass term $ z^{2} $ and/or fermion bilinears 
$ \eta_1 \eta_2 + \cdots + \eta_{2m-1} \eta_{2m} $ so that
\begin{equation}
\label{chatCondition}
(n + 1) - 2m - 2 = \hat{c} \, .
\end{equation}}     
\end{enumerate}
We thus obtain the modified superpotential
\begin{equation}
\widetilde{W} = \left\{
  \begin{array}{l}
     W  \quad  ( \hat{c} = N - 2 ) \\
     W + z^{2} \quad ( \hat{c} > N - 2 ) \\
     W + \sum_{k=1}^{m} \eta_{2k-1} \eta_{2k}
     \quad ( \hat{c} < N - 2, \ \sum_{a=1}^N q_{z_a} \textrm{integral} ) 
     \\
     W +  z^2 +  \sum_{k=1}^{m} \eta_{2k-1} \eta_{2k}
     \quad ( \hat{c} < N - 2, \ \sum_{a=1}^N q_{z_a} 
             \textrm{half-integral} ) \, .
  \end{array}
\right.
\end{equation}
The added fields have no effect on the chiral ring or the conformal
fixed point to which the theory flows.
The condition (\ref{chatCondition}) allows a change of variables
with constant Jacobian to be made such that one of the new variables
appears only linearly in the modified superpotential.
The change of variables is not one-to-one, so the modified theory must
orbifolded by the diagonal subgroup of its phase symmetries.
When $ m = 0 \ ( m \neq 0 ) $, integrating the linear new 
variable out of the path integral for the modified action as a Lagrange 
multiplier yields a (super-)delta function constraint which corresponds 
to the bosonic variety $ \mathcal{X} $ (supervariety $ \mathcal{M} $)
defined by $ \widetilde{W} = 0 $ having vanishing (super-)first Chern 
class.
The first Chern class of $ \mathcal{X} $ vanishes when
\begin{equation}
   \sum_{\mu=1}^{n+1} n_{z_{\mu}} - d = 0 \, ,
\end{equation}
whereas the super-first Chern class of $ \mathcal{M} $ vanishes when
\begin{equation}
\label{super-firstChern}
   \sum_{\mu=1}^{n+1} n_{z_{\mu}}
 - \sum_{\alpha=1}^{2m} n_{ \eta_{\alpha} } - d = 0 \, .
\end{equation}

\section{\label{Fermionic}Fermionic weights and the singular locus 
constraint}

As described in the Introduction, we require the fermionic weights 
to be consistent with the singular locus constraint
(\ref{SingularLocusConstraint}) and the quasihomogeneity constraint 
(\ref{QuasihomogeneityConstraint}).
The restriction placed on the fermionic weights by the latter
constraint is obvious.
Let us now explain what consistency with the former constraint means.

The supervariety $ \mathcal{M} $ defined by the hypersurface 
$ \widetilde{W} = 0 $ has a $ \mathbf{Z}_p $ fixed point set under the 
weighted projective identification 
(\ref{WeightedProjectiveIdentification}) if and only if the following 
conditions are both satisfied:
\begin{enumerate}
\item{The index set 
$ B_p \equiv \{ \mu | \lambda^{ n_{z_{\mu}} }_p = 1 \} $
is nonempty, where $ \lambda_p \equiv e^{2 \pi i / p} $.}
\item{The quantity
\begin{equation}
\label{D^bos_p}
\mathcal{D}_p
 = \left\{
          \begin{array}{l}
             | B_p | - 2 \quad (d/p \in \mathbf{Z})  \\
             | B_p | - 1 \quad (d/p \notin \mathbf{Z}) \, .
          \end{array}
    \right.
\end{equation}
satisfies $ \mathcal{D}_p \geq 0 $ if the index set
$ F_p \equiv \{ \alpha | \lambda^{ n_{\eta_{\alpha}} }_p = 1 \} $ is 
empty.
}
\end{enumerate} 
There are no purely fermionic $ \mathbf{Z}_p $ fixed point sets because 
the body of $ \mathcal{M} $ is embedded in $ \mathbf{WP}^n $.
The superdimension of the $ \mathbf{Z}_p $ fixed point sets which do 
exist is given by
\begin{equation}
\label{Dtilde_p}
\widetilde{\mathcal{D}}_p
 = \left\{
          \begin{array}{l}
             | B_p | - | F_p | - 2 \quad (d/p \in \mathbf{Z})  \\
             | B_p | - | F_p | - 1 \quad (d/p \notin \mathbf{Z}) \, .
          \end{array}
    \right.
\end{equation}
In the case $ d/p \in \mathbf{Z} $, the $ -2 $ arises because
the weighted projective identification and the hypersurface equation
each reduce the superdimension by 1.
In the case $ d/p \notin \mathbf{Z} $, the hypersurface equation
is identically satisfied and hence does not reduce the superdimension.
Consistency with the singular locus constraint
(\ref{SingularLocusConstraint}) means that
\begin{equation}
\widetilde{\mathcal{D}}_p \leq \hat{c} - 1 \quad \forall p \, ,
\end{equation}
where we have used (\ref{superdimension}).
We see from (\ref{Dtilde_p}) that $ \mathbf{Z}_p $ fixed point sets with 
$ | B_p | \leq | F_p | $ have 
negative superdimension and hence are consistent with  
(\ref{SingularLocusConstraint}).
Thus, when checking for consistency with 
(\ref{SingularLocusConstraint}), we can focus our attention on $ 
\mathbf{Z}_p $ fixed point sets with $ | B_p | > | F_p | $.
Note that the bosonic part of such a $ \mathbf{Z}_p $ fixed point set 
has complex dimension given by (\ref{D^bos_p}).

To illustrate the above, let us consider a concrete example.
\newtheorem{Exa}{Example}[section]
\begin{Exa}
\label{Example3.1}
Consider a Gepner model with level/invariant structure  
$ 10_D $  $ 10_D $  $ 1_A $  $ 1_A $  $ 1_A $  $ 1_A $.
Following the procedure described in Section \ref{Gepner/NLSM}, we 
obtain the quasihomogeneous degree $ d = 12 $ modified superpotential
\begin{equation*}
\widetilde{W} =   \sum_{i=1}^{2} (x^6_i + x_i y^2_i) 
                + \sum_{i=3}^{6} x^3_i 
                + z^{2} + \eta_1 \eta_2 + \eta_3 \eta_4 \, .
\end{equation*}
The hypersurface $ \widetilde{W} = 0 $ defines a supervariety 
$ \mathcal{M} $ embedded in 
$$ 
\mathbf{WSP}(2,5,2,5,4,4,4,4,6| 
     n_{\eta_1},n_{\eta_2},n_{\eta_3},n_{\eta_4}) \, . 
$$
We will denote the family of quasihomogeneous degree $ d = 12 $ 
hypersurfaces embedded in this weighted superprojective space by
$$
\mathbf{WSP}(2,5,2,5,4,4,4,4,6|
             n_{\eta_1},n_{\eta_2},n_{\eta_3},n_{\eta_4})[12] \, .
$$
A hypersurface in this family would also be obtained from a Gepner model 
with level and invariant structure 
$ 4_D $  $ 4_D $  $ 10_D $  $ 10_D $ or
$ 4_D $  $ 10_D $ $ 10_D $  $  1_A $ $ 1_A $.

According to (\ref{QuasihomogeneityConstraint}), the fermions in each 
bilinear can take on the values (modulo a relabelling of the fermions)
\begin{equation}
( n_{ \eta_{2k-1} }, n_{ \eta_{2k} } ) 
  \in \{ (1,11), (2,10), (3,9), (4,8), (5,7), (6,6) \} \, .
\end{equation}
To further constrain the fermionic weights, we now consider the 
$ \mathbf{Z}_p \ (p = 2,4,5) $ fixed point sets.
First, we determine the complex dimension of the bosonic parts
of these fixed point sets:

\begin{itemize}

\item[]{$ \underline{p = 2} $: 
The bosonic part of the $ \mathbf{Z}_2 $ fixed point set is
\begin{equation*}
\sum_{i=1}^2 (x^6_i + x_i y^2_i) + \sum_{i=3}^6 x^3_i + z^2 = 0 \, ,
\quad
y_1 = y_2 = 0 \, .
\end{equation*}
There are $ |B_2| = 7 $ bosons 
$ ( x_i \ (i=1,\ldots,6) \ \mathrm{and} \ z ) $ 
in this fixed point set.
Since $ d/p = 6 \in \mathbf{Z} $, (\ref{D^bos_p}) gives
\begin{equation} 
\mathcal{D}_2 = | B_2 | - 2 = 5 \, .
\end{equation}
}
\item[]{ $ \underline{p = 4} $:
The bosonic part of the $ \mathbf{Z}_4 $ fixed point set is
\begin{equation*}
\sum_{i=3}^6 x^3_i = 0 \, ,
\quad
x_i = y_i = 0 \quad (i = 1,2) \, ,
\quad
z = 0 \, .
\end{equation*}
There are $ | B_4 | = 4 $ bosons 
$ ( x_i \ (i=3,\ldots,6) ) $ in this fixed point set.
Since $ d/p = 3 \in \mathbf{Z} $, (\ref{D^bos_p}) gives
\begin{equation}
\mathcal{D}_4 = | B_4 | - 2 = 2 \, .
\end{equation}
}
\item[]{$ \underline{p = 5} $: 
The bosonic part of the $ \mathbf{Z}_5 $ fixed point set is
\begin{equation*}
\sum_{i=1}^2 (x^6_i + x_i y^2_i) = 0 \, ,
\quad
x_i  = 0 \quad (i = 1,\ldots,6) \, ,
\quad
z = 0 \, .
\end{equation*}
There are $ | B_5 | = 2 $ bosons
$ ( y_1 \ \mathrm{and} \ y_2 ) $
in this fixed point set.
Since $ d/p = 12/5 \notin \mathbf{Z} $, (\ref{D^bos_p}) gives
\begin{equation}
\mathcal{D}_5 = | B_5 | - 1 = 1 \, .
\end{equation}
}
\end{itemize}
Next, let 
$ \widetilde{\mathcal{D}}^{(j)}_p 
   = \mathcal{D}_p - | F^{(j)}_p | $ 
be the superdimension of the $ \mathbf{Z}_p $ fixed point set for the $ 
j $th distinct (modulo relabelling) fermionic weight combination
$ ( n_{\eta_1}, n_{\eta_2},  n_{\eta_3}, n_{\eta_4} )_j $
consistent with (\ref{QuasihomogeneityConstraint}).
For $ m=2 $ fermion bilinears, there are 
$ \frac{1}{2} \left[ \frac{d}{2} \right]
              \left( \left[ \frac{d}{2} \right] + 1 \right) $
such combinations, where $ [x] $ denotes the integer part of $ x $.  
Thus, for this example, $ j = 1,\ldots,21 $.
The $ \widetilde{\mathcal{D}}^{(j)}_p $ are given in Table \ref{D^j_pTable}.
\TABLE{
\label{D^j_pTable}
\centerline{
\begin{tabular}{rcccc}
$ j $ & $ ( n_{\eta_1}, n_{\eta_2},  n_{\eta_3}, n_{\eta_4} )_j $
& $ \widetilde{\mathcal{D}}^{(j)}_2 $
& $ \widetilde{\mathcal{D}}^{(j)}_4 $
& $ \widetilde{\mathcal{D}}^{(j)}_5 $ \\ \hline
 1 & (1,11,1,11) & 5 & 2 & 1 \\
 2 & (1,11,2,10) & 3 & 2 & 0 \\
 3 & (1,11,3,9) & 5 & 2 & 1 \\
 4 & (1,11,4,8) & 3 & 0 & 1 \\
 5 & (1,11,5,7) & 5 & 2 & 0 \\
 6 & (1,11,6,6) & 3 & 2 & 1 \\
 7 & (2,10,2,10) & 1 & 2 & -1 \\
 8 & (2,10,3,9) & 3 & 2 & 0 \\
 9 & (2,10,4,8) & 1 & 0 & 0 \\
10 & (2,10,5,7) & 3 & 2 & -1 \\
11 & (2,10,6,6) & 1 & 2 & 0 \\
12 & (3,9,3,9) & 5 & 2 & 1 \\
13 & (3,9,4,8) & 3 & 0 & 1 \\
14 & (3,9,5,7) & 5 & 2 & 0 \\
15 & (3,9,6,6) & 3 & 2 & 1 \\
16 & (4,8,4,8) & 1 & -2 & 1 \\
17 & (4,8,5,7) & 3 & 0 & 0 \\
18 & (4,8,6,6) & 1 & 0 & 1 \\
19 & (5,7,5,7) & 5 & 2 & -1\\
20 & (5,7,6,6) & 3 & 2 & 0 \\
21 & (6,6,6,6) & 1 & 2 & 1
\end{tabular}
\caption{Computation of 
$ \widetilde{\mathcal{D}}^{(j)}_p 
   = \mathcal{D}_p - | F^{(j)}_p |  \ (p=2,4,5) $.}}}
Finally, exclude all fermionic weight combinations which do not satisfy
$$ 
\widetilde{\mathcal{D}}^{(j)}_p \leq \hat{c} - 1 = 2 \ \forall p \, . 
$$
This leaves the fermionic weight combinations 
\begin{equation*}
\begin{array}{c}
(2,10,2,10) \, , \ (2,10,4,8) \, , \ (2,10,6,6) \, , 
\ (4,8,4,8) \, , \ (4,8,6,6) \, , \ (6,6,6,6) \, . 
\end{array}
\end{equation*}
We can use the parameters $ k $ and $ l $ defined in the Appendix 
to express these combinations in the compact form 
$ (2k,12-2k,2l,12-2l) $.
Here, $ k = 1,\ldots,3 $ and $ l = 1,\ldots,3 $.
It is understood that repeated fermionic weight combinations 
generated with this notation are counted only once.
In this notation, the hypersurface 
$ \widetilde{W} = 0 $ defines a supervariety $ \mathcal{M} $ 
embedded in
$ \mathbf{WSP}(2,5,2,5,4,4,4,4,6|2k,12-2k,2l,12-2l) $ 
and is a member of hypersurface family 250 of the Appendix:
\begin{equation*}  
\mathbf{WSP}(2,5,2,5,4,4,4,4,6|2k,12-2k,2l,12-2l)[12] \, .
\end{equation*}
\end{Exa}

\section{\label{Analysis}Analysis}

We have written a computer program which takes as input data which 
encodes the superpotential $ W $ for $ \hat{c} = 3 $ Gepner models.
Next, the program determines the modified superpotential 
$ \widetilde{W} $ as explained in Section \ref{Gepner/NLSM}.
When $ \widetilde{W} $ depends on five bosonic fields and no fermionic 
fields, the output of the program is the hypersurface family
$$ \mathbf{WP}(n_{z_1},\ldots,n_{z_{n+1}})[d] $$ 
corresponding to the hypersurface $ \widetilde{W} = 0 $ which defines a 
bosonic variety $ \mathcal{X} $ embedded in 
$ \mathbf{WP}(n_{z_1},\ldots,n_{z_{n+1}}) $.
For the remaining cases, the output is the hypersurface family
$$
\mathbf{WSP}(n_{z_1},\ldots,n_{z_{n+1}}|
             n_{\eta_1},\ldots,n_{\eta_{2m}})[d]
$$ 
corresponding to the hypersurface 
$ \widetilde{W} = 0 $ which defines a supervariety 
$ \mathcal{M} $ embedded in 
$ \mathbf{WSP}(n_{z_1},\ldots,n_{z_{n+1}} |n_{\eta_1}, 
\ldots,n_{\eta_{2m}}) $. 
The fermionic weights are determined by requiring the singular locus
constraint (\ref{SingularLocusConstraint}) and the quasihomogeneity 
constraint (\ref{QuasihomogeneityConstraint}) to be satisfied. 
In this manner, we obtain at least one fermionic weight solution
for each $ \hat{c} = 3 $ Gepner model corresponding to a supervariety 
$ \mathcal{M} $ through Sethi's proposed correspondence.
The 254 hypersurface families associated with these supervarieties are
tabulated in the Appendix.
When we replace the singular locus constraint 
(\ref{SingularLocusConstraint}) with Conjecture 
\ref{ConjecturedSingularLocusConstraint}, this results in models 
corresponding to the hypersurface families 50, 94, 95, 121, 125, and 229 
of the Appendix having no solution for the fermionic weights.

In principle, the singularities which do arise could be determined by 
identifying the fermionic weights which yield agreement between the 
Hodge diamond of the Landau-Ginzburg orbifold and the Hodge diamond of 
$ \mathcal{M} $.
The Landau-Ginzburg orbifold Hodge diamond can be computed using the 
techniques of \cite{LGOHodge}.
Such calculations can be done quickly with the software package PALP
\cite{KreuzerSkarke:PALP}.
Unfortunately, at present, there is no supercohomology theory which 
allows the Hodge diamond of $ \mathcal{M} $ to be computed.

The Hodge diamond of $ \mathcal{X} $ can be computed using the orbifold
techniques of \cite{Zaslow:Topological}.
This is possible because, with a change of coordinates
\begin{equation}
\label{change}
z_{\mu} = (\zeta_{\mu})^{ n_{z_{\mu}} } \, ,
\end{equation}
the hypersurface 
$ \mathbf{WP}(n_{z_1},\ldots,n_{z_{n+1}})[d] $ which defines 
$ \mathcal{X} $ can be written as an orbifold of a hypersurface in a 
homogeneous projective space $ \mathbf{P}^n $, i.e.
\begin{equation}
\mathbf{WP}(n_{z_1},\ldots,n_{ z_{n+1} })[d]
  = \frac{ \mathbf{P}^n[d] }
         {  \mathbf{Z}_{ n_{z_1} } \times \cdots \times 
            \mathbf{Z}_{ n_{z_{n+1}} }  } \, . 
\end{equation}
In contrast, the hypersurface 
$ \mathbf{WSP}(n_{z_1},\ldots,n_{z_{n+1}}|
               n_{\eta_1},\ldots,n_{\eta_{2m}})[d] $
which defines $ \mathcal{M} $ cannot be written as an orbifold of a 
hypersurface in homogeneous superprojective space 
$ \mathbf{SP}^{(n|2m)} $.
This is because the analogue of (\ref{change}) does not make sense for 
Grassmann coordinates. 
Nevertheless, as described in \cite{Sethi}, we can use orbifold 
considerations to gain insight into the structure of the Hodge diamond 
of $ \mathcal{M} $.
In the following examples, we will use this heuristic reasoning and 
compare the resulting fermionic weights with those obtained from the 
singular locus constraint (\ref{SingularLocusConstraint}) and 
Conjecture \ref{ConjecturedSingularLocusConstraint}.

\begin{Exa}
\label{Example4.1}
A hypersurface in the family
$ \mathbf{WSP}(1,1,2,2,2,2,2|n_{\eta_1}, n_{\eta_2})[6] $
can be obtained from a Gepner model with any of the following 
level/invariant structures:
\begin{align*}
4_D \ 4_D \ 4_A \ 4_A \ 1_A \, ,
\quad
4_D \ 4_A \ 4_A \ 1_A \ 1_A \ 1_A \, ,
\quad
4_A \ 4_A \ 1_A \ 1_A \ 1_A \ 1_A \ 1_A \, .
\end{align*}
For definiteness, we will focus on the last of these.
Following the procedure described in Section \ref{Gepner/NLSM}, we
obtain the modified superpotential
\begin{equation*}
\widetilde{W}
  =   x_1^{6} + x_2^{6} + x_3^3 + x_4^3 + x_5^3  + x_6^3 + x_7^3
    + \eta_1 \eta_2 \, .
\end{equation*}
This example was discussed in \cite{Sethi}.
Here, we simply note that cohomology considerations, Conjecture 
\ref{ConjecturedSingularLocusConstraint}, and, as indicated by 
hypersurface family 179 of the Appendix, the singular locus constraint 
(\ref{SingularLocusConstraint}) all yield the same unique result
$ (n_{\eta_1},n_{\eta_2}) = (2,4) $. 
\end{Exa}

\begin{Exa}
\label{Example4.2}
A hypersurface in the family
$ \mathbf{WSP}(1,1,4,4,4,4,6|n_{\eta_1}, n_{\eta_2})[12] $
can be obtained from a Gepner model with 
any of the following level/invariant structures:
\begin{gather*} 
4_D \ 10_A \ 4_A \ 2_A \ 1_A \, ,
\quad
10_A \ 4_A \ 2_A \ 1_A \ 1_A \ 1_A \, ,
\quad
4_D \ 4_D \ 10_A \ 10_A \, ,
\\
4_D \ 10_A \ 10_A \ 1_A \ 1_A \, ,
\quad
10_A \ 10_A \ 1_A \ 1_A \ 1_A \ 1_A \, .  
\end{gather*}
For definiteness, we will focus on the last of these.
Following the procedure described in Section \ref{Gepner/NLSM}, we 
obtain the modified superpotential
\begin{equation*}
\widetilde{W} 
  =   x_1^{12} + x_2^{12} + x_3^3 + x_4^3 + x_5^3  + x_6^3 
    + z^2 + \eta_1 \eta_2 \, .
\end{equation*}
Employing the heuristic reasoning of \cite{Sethi}, we find for 
$ (n_{\eta_1},n_{\eta_2}) = (4,8) $ that the Hodge diamond of 
$ \mathcal{M} $ is
\begin{align}
\label{DiamondExample4.2}
\begin{matrix}
& & & 1 & & & \\
& & 0 & & 0 & & \\
& 0 & & 7 & & 0 & \\
1 & & 79 & & 79 & & 1 \\  
& 0 & & 7 & & 0 & \\
& & 0 & & 0 & & \\
& & & 1 & & & \\
\end{matrix}
\quad
&=
\quad
\begin{matrix}
& & & 1 & & & \\
& & 0 & & 0 & & \\
& 0 & & 0 & & 0 & \\
1 & & 79 & & 79 & & 1 \\
& 0 & & 0 & & 0 & \\
& & 0 & & 0 & & \\
& & & 1 & & & \\
\end{matrix}
\quad
+
\quad
\begin{matrix}
& & & \Gray{0} \Black & & & \\
& & \Gray{0} \Black & & \Gray{0} \Black & & \\
& \Gray{0} \Black & & 6 & & \Gray{0} \Black & \\
\Gray{0} \Black & & \Gray{0} \Black & & \Gray{0} \Black & & \Gray{0} 
\Black \\
& \Gray{0} \Black & & 6 & & \Gray{0} \Black & \\
& & \Gray{0} \Black & & \Gray{0} \Black & & \\
& & & \Gray{0} \Black & & & \\
\end{matrix}
\quad 
+ 
\quad
\begin{matrix}
& & & \Gray{0} \Black & & & \\
& & \Gray{0} \Black & & \Gray{0} \Black & & \\
& \Gray{0} \Black & & 1 & & \Gray{0} \Black & \\
\Gray{0} \Black & & 0 & & 0 & & \Gray{0} \Black \\
& \Gray{0} \Black & & 1 & & \Gray{0} \Black & \\
& & \Gray{0} \Black & & \Gray{0} \Black & & \\
& & & \Gray{0} \Black & & & \\
\end{matrix} \, .
\end{align}
The first term on the right-hand side of (\ref{DiamondExample4.2}) is 
the contribution arising from the untwisted sector.
The second term includes the contribution of the fixed point set 
associated with the third twisted sector (the upper 6) and the 
fixed point set associated with the ninth twisted sector (the lower 6).
Finally, the last term arises from the identity and volume forms of the 
fixed point set associated with the sixth twisted sector.
Our result (\ref{DiamondExample4.2}) agrees with the Hodge diamond of 
the associated Landau-Ginzburg orbifold. 
For all other fermionic weight assigments consistent with
(\ref{QuasihomogeneityConstraint}), we do not obtain this agreement.
Thus, in this example, using these heuristic arguments, we obtain a 
unique result 
$$ (n_{\eta_1},n_{\eta_2}) = (4,8) \, . $$ 
We note that this result agrees with what would be obtained from 
Conjecture \ref{ConjecturedSingularLocusConstraint}.    
In contrast, as indicated by hypersurface family 10 of the Appendix, the 
singular locus constraint (\ref{SingularLocusConstraint}) allows 
$$ (n_{\eta_1},n_{\eta_2}) \in \{ (2,10), (4,8), (6,6) \} \, . $$
\end{Exa}

\begin{Exa}
\label{Example4.3}
A hypersurface in the family
$ \mathbf{WSP}(1,3,3,3,4,4,6|n_{\eta_1}, n_{\eta_2})[12] $
can be obtained from a Gepner model with
either of the following level/invariant structures:
\begin{equation*}
4_D \ 10_A \ 2_A \ 2_A \ 2_A \, ,
\quad
10_A \ 2_A \ 2_A \ 2_A \ 1_A \ 1_A \, .
\end{equation*}
For definiteness, we will focus on the last of these.
Following the procedure described in Section \ref{Gepner/NLSM}, we
obtain the modified superpotential
\begin{equation*}
\widetilde{W} = x_1^{12} + x_2^4 + x_3^4 + x_4^4 + x_5^3 + x_6^3 + z^2 + 
                \eta_1 \eta_2 \,.
\end{equation*}
Employing the heuristic reasoning of \cite{Sethi}, we find for 
$ (n_{\eta_1},n_{\eta_2}) = (3,9) $ that the Hodge diamond of 
$ \mathcal{M} $ is 
\begin{align}
\label{DiamondExample4.3}
\begin{matrix}
& & & 1 & & & \\
& & 0 & & 0 & & \\
& 0 & & 10 & & 0 & \\
1 & & 46 & & 46 & & 1 \\
& 0 & & 10 & & 0 & \\
& & 0 & & 0 & & \\
& & & 1 & & & \\
\end{matrix}
\quad
&=
\quad
\begin{matrix}
& & & 1 & & & \\
& & 0 & & 0 & & \\
& 0 & & 0 & & 0 & \\
1 & & 46 & & 46 & & 1 \\
& 0 & & 0 & & 0 & \\
& & 0 & & 0 & & \\
& & & 1 & & & \\
\end{matrix}
\quad
+
\quad
\begin{matrix}
& & & \Gray{0} \Black & & & \\
& & \Gray{0} \Black & & \Gray{0} \Black & & \\
& \Gray{0} \Black & & 2 & & \Gray{0} \Black & \\
\Gray{0} \Black & & \Gray{0} \Black & & \Gray{0} \Black & & \Gray{0} 
\Black \\
& \Gray{0} \Black & & 2 & & \Gray{0} \Black & \\
& & \Gray{0} \Black & & \Gray{0} \Black & & \\
& & & \Gray{0} \Black & & & \\
\end{matrix}
\quad
+
\quad
\begin{matrix}
& & & \Gray{0} \Black & & & \\
& & \Gray{0} \Black & & \Gray{0} \Black & & \\
& \Gray{0} \Black & & 1 & & \Gray{0} \Black & \\
\Gray{0} \Black & & 0 & & 0 & & \Gray{0} \Black \\
& \Gray{0} \Black & & 1 & & \Gray{0} \Black & \\
& & \Gray{0} \Black & & \Gray{0} \Black & & \\
& & & \Gray{0} \Black & & & \\
\end{matrix}
\nonumber
\\
&\quad 
+
\quad
\begin{matrix}
& & & \Gray{0} \Black & & & \\
& & \Gray{0} \Black & & \Gray{0} \Black & & \\
& \Gray{0} \Black & & 7 & & \Gray{0} \Black & \\
\Gray{0} \Black & & \Gray{0} \Black & & \Gray{0} \Black & & \Gray{0} 
\Black \\
& \Gray{0} \Black & & 7 & & \Gray{0} \Black & \\
& & \Gray{0} \Black & & \Gray{0} \Black & & \\
& & & \Gray{0} \Black & & & \\
\end{matrix} \, .
\end{align}
The first three terms on the right-hand side of 
(\ref{DiamondExample4.3}) have the same origin as the corresponding 
terms in Example \ref{Example4.2}.
The fourth term includes the contribution of the fixed point set
associated with the fourth twisted sector (the lower 7) and the
fixed point set associated with the ninth twisted sector (the upper 7).
Our result (\ref{DiamondExample4.3}) agrees with the Hodge diamond of 
the associated Landau-Ginzburg orbifold.
For all other fermionic weight assigments consistent with
(\ref{QuasihomogeneityConstraint}), we do not obtain this agreement.
Thus, in this example, using these heuristic arguments, we obtain a
unique result 
$$ (n_{\eta_1},n_{\eta_2}) = (3,9) \, . $$ 
In contrast, Conjecture \ref{ConjecturedSingularLocusConstraint} allows 
$$ (n_{\eta_1},n_{\eta_2}) \in \{ (3,9), (6,6) \} $$ 
and, as indicated by hypersurface family 8 of the Appendix, the
singular locus constraint (\ref{SingularLocusConstraint}) allows
$$ 
(n_{\eta_1},n_{\eta_2}) 
   \in \{ (1,11), (2,10), (3,9), (4,8), (5,7), (6,6) \} \, . 
$$
\end{Exa}

\begin{Exa}
\label{Example4.4}
A hypersurface in the family
$ \mathbf{WSP}(2,3,1,2,2,2,4|n_{\eta_1}, n_{\eta_2})[8] $
can be obtained from a Gepner model with level/invariant structure
\begin{gather*}
6_D \ 6_A \ 2_A \ 2_A \ 2_A \, .
\end{gather*}
Following the procedure described in Section \ref{Gepner/NLSM}, we
obtain the modified superpotential
\begin{equation*}
\widetilde{W}
  =   x_1^{4} + x_1 y_1^2 + x_2^{8} + x_3^4 + x_4^4 + x_5^4
    + z^2 + \eta_1 \eta_2 \, .
\end{equation*}
Employing the heuristic reasoning of \cite{Sethi}, we find for
$ (n_{\eta_1},n_{\eta_2}) = (4,4) $ that the Hodge diamond of
$ \mathcal{M} $ is
\begin{align}
\label{DiamondExample4.4}
\begin{matrix}
& & & 1 & & & \\
& & 0 & & 0 & & \\
& 0 & & 1 & & 0 & \\
1 & & 73 & & 73 & & 1 \\
& 0 & & 1 & & 0 & \\
& & 0 & & 0 & & \\
& & & 1 & & & \\
\end{matrix}
\quad
&=
\quad
\begin{matrix}
& & & 1 & & & \\
& & 0 & & 0 & & \\
& 0 & & 0 & & 0 & \\
1 & & 63 & & 63 & & 1 \\
& 0 & & 0 & & 0 & \\
& & 0 & & 0 & & \\
& & & 1 & & & \\
\end{matrix}
\quad
+
\quad
\begin{matrix}
& & & \Gray{0} \Black & & & \\
& & \Gray{0} \Black & & \Gray{0} \Black & & \\
& \Gray{0} \Black & & 1 & & \Gray{0} \Black & \\
\Gray{0} \Black & & 10 & & 10 & & \Gray{0} \Black \\
& \Gray{0} \Black & & 1 & & \Gray{0} \Black & \\
& & \Gray{0} \Black & & \Gray{0} \Black & & \\
& & & \Gray{0} \Black & & & \\
\end{matrix} \, .
\end{align}
The first term on the right-hand side of (\ref{DiamondExample4.4}) is 
the contribution arising from the untwisted sector whereas the second 
term arises from the fourth twisted sector.
Our result (\ref{DiamondExample4.4}) agrees with the Hodge diamond of
the associated Landau-Ginzburg orbifold.
For all other fermionic weight assigments consistent with
(\ref{QuasihomogeneityConstraint}), we do not obtain this agreement.
Thus, in this example, using these heuristic arguments, we obtain a 
unique result
$$
(n_{\eta_1},n_{\eta_2}) = (4,4) \, .
$$
In contrast, Conjecture \ref{ConjecturedSingularLocusConstraint} and, as 
indicated by hypersurface family 3 of the Appendix, the singular locus 
constraint (\ref{SingularLocusConstraint}) both allow
$$
(n_{\eta_1},n_{\eta_2}) \in \{ (2,6), (4,4) \} \, .
$$
\end{Exa}

\begin{Exa}
\label{Example4.5}
Let us revisit the family of hypersurfaces discussed in Example 
\ref{Example3.1},
$
\mathbf{WSP}(2,5,2,5,4,4,4,4,6|
             n_{\eta_1},n_{\eta_2},n_{\eta_3},n_{\eta_4})[12] 
$.
We again focus on the hypersurface obtained from a Gepner 
model with level/invariant structure 
$ 10_D $  $ 10_D $  $ 1_A $  $ 1_A $  $ 1_A $  $ 1_A $.
Proceeding as in the above examples, we find for
$ (n_{\eta_1},n_{\eta_2},n_{\eta_3},n_{\eta_4})= (2,10,4,8) $ that the 
Hodge diamond of $ \mathcal{M} $ is
\begin{align}
\label{DiamondExample4.5}
\begin{matrix}
& & & 1 & & & \\
& & 0 & & 0 & & \\
& 0 & & 7 & & 0 & \\
1 & & 79 & & 79 & & 1 \\
& 0 & & 7 & & 0 & \\
& & 0 & & 0 & & \\
& & & 1 & & & \\
\end{matrix}
\quad
&=
\quad
\begin{matrix}
& & & 1 & & & \\
& & 0 & & 0 & & \\
& 0 & & 0 & & 0 & \\
1 & & 47 & & 47 & & 1 \\
& 0 & & 0 & & 0 & \\
& & 0 & & 0 & & \\
& & & 1 & & & \\
\end{matrix}
\quad
+
\quad
\begin{matrix}
& & & \Gray{0} \Black & & & \\
& & \Gray{0} \Black & & \Gray{0} \Black & & \\
& \Gray{0} \Black & & 6 & & \Gray{0} \Black & \\
\Gray{0} \Black & & \Gray{0} \Black & & \Gray{0} \Black & & \Gray{0} 
\Black \\
& \Gray{0} \Black & & 6 & & \Gray{0} \Black & \\
& & \Gray{0} \Black & & \Gray{0} \Black & & \\
& & & \Gray{0} \Black & & & \\
\end{matrix}
\quad
+
\quad
\begin{matrix}
& & & \Gray{0} \Black & & & \\
& & \Gray{0} \Black & & \Gray{0} \Black & & \\
& \Gray{0} \Black & & 1 & & \Gray{0} \Black & \\
\Gray{0} \Black & & 32 & & 32 & & \Gray{0} \Black \\
& \Gray{0} \Black & & 1 & & \Gray{0} \Black & \\
& & \Gray{0} \Black & & \Gray{0} \Black & & \\
& & & \Gray{0} \Black & & & \\
\end{matrix} \, .
\end{align}
The terms on the right-hand side of (\ref{DiamondExample4.5}) originate 
from the same sectors as the corresponding terms in Example 
\ref{Example4.2}.
Our result (\ref{DiamondExample4.5}) agrees with the Hodge diamond of 
the associated Landau-Ginzburg orbifold.
For all other fermionic weight assigments consistent with
(\ref{QuasihomogeneityConstraint}), we do not obtain this agreement.
Thus, in this example, using these heuristic arguments, we obtain
$$ (n_{\eta_1},n_{\eta_2},n_{\eta_3},n_{\eta_4}) 
   \in \{ (2,10,4,8), (4,8,6,6) \} \, . $$ 
In contrast, Conjecture \ref{ConjecturedSingularLocusConstraint} allows  
$$ 
(n_{\eta_1},n_{\eta_2},n_{\eta_3},n_{\eta_4}) 
   \in \{ (2,10,4,8), (4,8,4,8), (4,8,6,6) \} 
$$
and the singular locus constraint (\ref{SingularLocusConstraint}) allows
\begin{align*}
(n_{\eta_1},n_{\eta_2},n_{\eta_3},n_{\eta_4}) 
   \in \{ 
               & (2,10,2,10), (2,10,4,8), (2,10,6,6),  
\\
               & (4,8,4,8), (4,8,6,6), (6,6,6,6)
       \} \, . 
\end{align*}
\end{Exa}
It is interesting to note that, in the above examples, the solutions 
obtained from the heuristic approach agree precisely with those obtained 
from Conjecture \ref{ConjecturedSingularLocusConstraint} supplemented by 
the constraint
\begin{equation}
\label{SupplementaryConstraint}
\widetilde{\mathcal{D}}_p \geq 0 \quad \textrm{whenever} \quad 
\mathcal{D}_p \geq 0 \, .
\end{equation}
Let us now consider an example in which the heuristic approach yields no 
solution.

\begin{Exa}
\label{Example4.6}
A hypersurface in the family
$ \mathbf{WSP}(1,1,1,2,2,2,3|n_{\eta_1}, n_{\eta_2})[6] $
can be obtained from a Gepner model with either of the following 
level/invariant structures:
\begin{equation*}
4_D \ 4_A \ 4_A \ 4_A \ 1_A \, ,
\quad
4_A \ 4_A \ 4_A \ 1_A \ 1_A \ 1_A \, .
\end{equation*}
For definiteness, we will focus on the last of these.
Following the procedure described in Section \ref{Gepner/NLSM}, we
obtain the modified superpotential
\begin{equation*}
\widetilde{W}
  =   x_1^{6} + x_2^{6} + x_3^{6} + x_4^3 + x_5^3 + x_6^3
    + z^2 + \eta_1 \eta_2 \, .
\end{equation*}
Employing the heuristic reasoning of \cite{Sethi}, we find for 
$ (n_{\eta_1},n_{\eta_2}) \in \{ (1,5), (3,3) \} $ that the Hodge 
diamond of $ \mathcal{M} $ is
\begin{align}
\label{DiamondExample4.6}
\begin{matrix}
& & & 1 & & & \\
& & 0 & & 0 & & \\
& 0 & & 1 & & 0 & \\
1 & & 84 & & 84 & & 1 \\
& 0 & & 1 & & 0 & \\
& & 0 & & 0 & & \\
& & & 1 & & & \\
\end{matrix}
\quad
&=
\quad
\begin{matrix}
& & & 1 & & & \\
& & 0 & & 0 & & \\
& 0 & & 0 & & 0 & \\
1 & & 83 & & 83 & & 1 \\
& 0 & & 0 & & 0 & \\
& & 0 & & 0 & & \\
& & & 1 & & & \\
\end{matrix}
\quad
+
\quad
\begin{matrix}
& & & \Gray{0} \Black & & & \\
& & \Gray{0} \Black & & \Gray{0} \Black & & \\
& \Gray{0} \Black & & 1 & & \Gray{0} \Black & \\
\Gray{0} \Black & & 1 & & 1 & & \Gray{0} \Black \\
& \Gray{0} \Black & & 1 & & \Gray{0} \Black & \\
& & \Gray{0} \Black & & \Gray{0} \Black & & \\
& & & \Gray{0} \Black & & & \\
\end{matrix} \, .
\end{align}
The first term on the right-hand side of 
(\ref{DiamondExample4.6}) is the contribution arising from the 
untwisted sector.
The second term includes the contribution of the fixed point set
associated with the third twisted sector.  
For $ (n_{\eta_1},n_{\eta_2}) = (2,4) $, we find that the only 
contribution to the Hodge diamond of $ \mathcal{M} $ arises from the 
untwisted sector.
In all cases, the result disagrees with the Landau-Ginzburg orbifold 
Hodge diamond
\begin{equation}
\begin{matrix}
\label{DiamondExample4.6LGO}
& & & 1 & & & \\
& & 0 & & 0 & & \\
& 0 & & 0 & & 0 & \\
1 & & 84 & & 84 & & 1 \\
& 0 & & 0 & & 0 & \\
& & 0 & & 0 & & \\
& & & 1 & & & \\
\end{matrix} \, .
\end{equation}
Thus, in this example, using these heuristic arguments, we find no 
solution for the fermionic weights.
In contrast, Conjecture \ref{ConjecturedSingularLocusConstraint} 
and, as indicated by hypersurface family 2 of the Appendix, the 
singular locus constraint (\ref{SingularLocusConstraint}) both allow
$ (n_{\eta_1},n_{\eta_2}) \in \{ (1,5), (2,4), (3,3) \}. $ 

\end{Exa}

\section{\label{Conclusion}Conclusion}

The analysis in Section \ref{Analysis} compares the fermionic weights 
obtained from our computer program with those obtained 
by requiring agreement between the Hodge diamond of the Landau-Ginzburg 
orbifold and the heuristically determined Hodge diamond of the 
supervariety.
Running the program with the singular locus constraint 
(\ref{SingularLocusConstraint}) yields at least one solution for each 
$ \hat{c} = 3 $ Gepner model associated with a supervariety through 
Sethi's proposed correspondence.
Conjecture \ref{ConjecturedSingularLocusConstraint} is a stronger 
constraint, but it still yields at least one solution for the vast 
majority of these models.
The heuristic approach places the strongest constraint on the 
fermionic weights.
It yields a unique solution in Examples 
\ref{Example4.1}, \ref{Example4.2}, 
\ref{Example4.3}, and \ref{Example4.4}, two solutions in Example 
\ref{Example4.5}, and no solution in Example \ref{Example4.6}.
In the examples we have studied, the heuristically determined solutions 
are a subset of the solutions obtained from Conjecture 
\ref{ConjecturedSingularLocusConstraint}.
Furthermore, in these examples, when the heuristic approach yields 
solutions, these solutions agree precisely
with those obtained from Conjecture 1.1 supplemented by the constraint
(\ref{SupplementaryConstraint}).
Thus, something seems to be ``right'' about the combination of 
Conjecture \ref{ConjecturedSingularLocusConstraint} and the constraint 
(\ref{SupplementaryConstraint}).
A proper supercohomology theory would allow more conclusive statements 
to be made.

In the Appendix, Table \ref{ResultsTable} indicates when the Newton 
polytope of $ \widetilde{W}_{bos} $ admits a nef partition.
In this case, the Landau-Ginzburg orbifold can be given a geometrical 
interpretation as a nonlinear sigma model on a complete intersection 
Calabi-Yau manifold.
The complete intersection Calabi-Yau manifold should be equivalent
to the Calabi-Yau supermanifold prescribed by Sethi's proposal.
It can be shown that a reflexive Gorenstein polytope $ \Delta $ admits a 
nef partition if and only if the reflexive Gorenstein cones 
$ \sigma_{\Delta} $ and $ \sigma^{\vee}_{\Delta} $ are both completely 
split \cite{nef}.
In fact, the Landau-Ginzburg orbifold can be given a complete 
intersection Calabi-Yau manifold interpretation even when only 
$ \sigma_{\Delta} $ is completely split \cite{nef}.
In Example \ref{Example4.4}, the Newton polytope of 
$ \widetilde{W}_{bos} $ is nonreflexive Gorenstein.
It turns out that, for all of the remaining examples in Section 
\ref{Analysis}, the Newton polytope of $ \widetilde{W}_{bos} $ is 
reflexive Gorenstein but $ \sigma_{\Delta} $ is not completely split.
Thus, in these examples, the Landau-Ginzburg orbifold cannot be given a 
complete intersection Calabi-Yau manifold interpretation.
We leave a detailed investigation of the cases in which only 
$ \sigma_{\Delta} $ or only $ \sigma^{\vee}_{\Delta} $
is completely split to future work.

\appendix
\section{\label{A}Supervariety hypersurface families}

In Table \ref{ResultsTable}, we list the supervariety hypersurface 
families associated with $ \hat{c} = 3 $ Gepner models.
A hypersurface family corresponding to a hypersurface
$ \widetilde{W} = 0 $ which defines a supervariety $ \mathcal{M} $
embedded in
$
\mathbf{WSP}(n_{z_1},\ldots,n_{z_{n+1}}|n_{\eta_1},\ldots,n_{\eta_{2m}})
$ is denoted by
$
\mathbf{WSP}(n_{z_1},\ldots,n_{z_{n+1}}|
             n_{\eta_1},\ldots,n_{\eta_{2m}})[d].
$
Here, $ \widetilde{W} $ is the modified superpotential obtained
by satisfying (\ref{chatCondition}) and $ d $ is its degree of
quasihomogeneity.
The fermionic weights are determined by requiring
(\ref{SingularLocusConstraint}) and
(\ref{QuasihomogeneityConstraint}) to be satisfied.
The solutions for these fermionic weights are parameterized by
$ k=1,\ldots,[ \frac{d}{2 s_k} ] $ and
$ l=1,\ldots,[ \frac{d}{2 s_l} ] $,
where $ s_k $ and $ s_l $ are the coefficients of $ k $ and $ l $ in
the first and third fermion weight assignments, respectively.
For each hypersurface family, the Hodge numbers
$ h^{1,1} $ and $ h^{2,1} $ and the Euler number
$ \chi = 2 \left( h^{1,1} - h^{2,1} \right) $ of the associated
Landau-Ginzburg orbifold are given.
When the Newton polytope of $ \widetilde{W}_{bos} $ admits a nef 
partition, this is indicated by ``nef''.
In a number of cases, the Landau-Ginzburg orbifold can be given a 
geometrical interpretation as a product of a two-torus and a K3 surface, 
which is indicated by ``$ T^2 \times K3 $''.
Finally, when the Newton polytope of $ \widetilde{W}_{bos} $ is 
nonreflexive Gorenstein, this is indicated by ``nonRG''.

\scriptsize
\begin{longtable}{rlrrrc}
\caption{Supervariety hypersurface families associated with 
$ \hat{c} = 3 $ Gepner models.} \\
\# &  hypersurface family & $ h^{1,1} $ & $ h^{2,1} $ & $ \chi $ & 
Comments \\ 
\endfirsthead
\# & hypersurface family & $ h^{1,1} $ & $ h^{2,1} $ & $ \chi $ & 
Comments \\
\endhead
1 & ${\bf WSP}(1,1,1,1,1,1,2|k,4-k)[4]$ & 0 & 90 & -180 \\ 
2 & ${\bf WSP}(1,1,1,2,2,2,3|k,6-k)[6]$ & 0 & 84 & -168 \\ 
3 & ${\bf WSP}(2,3,1,2,2,2,4|2k,8-2k)[8]$ & 1 & 73 & -144 & nonRG \\ 
4 & ${\bf WSP}(2,3,2,3,1,1,4|k,8-k)[8]$ & 1 & 77 & -152 \\ 
5 & ${\bf WSP}(2,4,2,4,1,2,5|2k,10-2k)[10]$ & 1 & 85 & -168 \\ 
6 & ${\bf WSP}(2,3,3,3,3,4,6|3k,12-3k)[12]$ & 1 & 61 & -120 \\ 
7 & ${\bf WSP}(2,2,3,3,4,4,6|2k,12-2k)[12]$ & 3 & 51 & -96 \\ 
8 & ${\bf WSP}(1,3,3,3,4,4,6|k,12-k)[12]$ & 10 & 46 & -72 \\
9 & ${\bf WSP}(1,2,3,4,4,4,6|2k,12-2k)[12]$ & 2 & 62 & -120 \\ 
10 & ${\bf WSP}(1,1,4,4,4,4,6|2k,12-2k)[12]$ & 7 & 79 & -144 \\ 
11 & ${\bf WSP}(2,5,2,3,3,3,6|k,12-k)[12]$ & 9 & 39 & -60 \\ 
12 & ${\bf WSP}(2,5,2,2,3,4,6|2k,12-2k)[12]$ & 2 & 74 & -144 & nonRG \\ 
13 & ${\bf WSP}(2,5,1,3,3,4,6|k,12-k)[12]$ & 1 & 61 & -120 & nonRG \\ 
14 & ${\bf WSP}(2,5,1,2,4,4,6|2k,12-2k)[12]$ & 3 & 75 & -144 & nonRG \\ 
15 & ${\bf WSP}(2,5,2,5,2,2,6|2k,12-2k)[12]$ & 2 & 128 & -252 \\ 
16 & ${\bf WSP}(2,5,2,5,1,3,6|k,12-k)[12]$ & 3 & 69 & -132 \\ 
17 & ${\bf WSP}(4,6,2,7,1,4,8|2k,16-2k)[16]$ & 3 & 75 & -144 \\ 
18 & ${\bf WSP}(2,7,2,7,2,4,8|2k,16-2k)[16]$ & 4 & 148 & -288 & nef \\ 
19 & ${\bf WSP}(6,4,2,3,6,6,9|6,12)[18]$ & 2 & 62 & -120 & nonRG \\ 
20 & ${\bf WSP}(6,4,6,4,1,6,9|2k,18-2k)[18]$ & 2 & 56 & -108 \\ 
21 & ${\bf WSP}(1,2,6,6,6,6,9|6,12)[18]$ & 8 & 68 & -120 \\ 
22 & ${\bf WSP}(2,8,2,3,6,6,9|2k,18-2k)[18]$ & 8 & 68 & -120 \\ 
23 & ${\bf WSP}(6,4,2,8,1,6,9|2k,18-2k)[18]$ & 4 & 76 & -144 \\ 
24 & ${\bf WSP}(2,8,2,8,1,6,9|2k,18-2k)[18]$ & 2 & 110 & -216 \\ 
25 & ${\bf WSP}(4,8,4,4,5,5,10|2k,20-2k)[20]$ & 21 & 21 & 0 & 
nef, $ T^2 \times K3 $ \\ 
26 & ${\bf WSP}(4,8,4,8,1,5,10|2k,20-2k)[20]$ & 13 & 49 & -72 \\ 
27 & ${\bf WSP}(2,9,4,5,5,5,10|k,20-k)[20]$ & 17 & 29 & -24 \\ 
28 & ${\bf WSP}(4,8,2,9,2,5,10|2k,20-2k)[20]$ & 7 & 79 & -144 & nonRG \\ 
29 & ${\bf WSP}(2,9,2,9,4,4,10|2k,20-2k)[20]$ & 7 & 143 & -272 & nef \\ 
30 & ${\bf WSP}(3,3,6,8,8,8,12|2k,24-2k)[24]$ & 21 & 21 & 0 & 
nef, $ T^2 \times K3 $ \\ 
31 & ${\bf WSP}(1,3,8,8,8,8,12|4k,24-4k)[24]$ & 16 & 52 & -72 \\ 
32 & ${\bf WSP}(6,9,3,4,6,8,12|6k,24-6k)[24]$ & 3 & 51 & -96 \\ 
33 & ${\bf WSP}(6,9,1,6,6,8,12|6k,24-6k)[24]$ & 10 & 46 & -72 & nonRG \\ 
34 & ${\bf WSP}(6,9,2,3,8,8,12|2k,24-2k)[24]$ & 13 & 37 & -48 & nonRG \\ 
35 & ${\bf WSP}(6,9,1,4,8,8,12|2k,24-2k)[24]$ & 12 & 48 & -72 & nonRG \\ 
36 & ${\bf WSP}(6,9,6,9,2,4,12|6k,24-6k)[24]$ & 6 & 66 & -120 & nef \\ 
37 & ${\bf WSP}(4,10,3,3,8,8,12|2k,24-2k)[24]$ & 11 & 35 & -48 \\ 
38 & ${\bf WSP}(6,9,4,10,3,4,12|2k,24-2k)[24]$ & 11 & 35 & -48 \\ 
39 & ${\bf WSP}(6,9,4,10,1,6,12|2k,24-2k)[24]$ & 7 & 55 & -96 & nonRG \\ 
40 & ${\bf WSP}(2,11,3,6,6,8,12|2k,24-2k)[24]$ & 10 & 46 & -72 \\ 
41 & ${\bf WSP}(2,11,3,4,8,8,12|2k,24-2k)[24]$ & 12 & 48 & -72 & nonRG 
\\ 
42 & ${\bf WSP}(2,11,1,6,8,8,12|2k,24-2k)[24]$ & 9 & 81 & -144 \\ 
43 & ${\bf WSP}(6,9,2,11,4,4,12|2k,24-2k)[24]$ & 6 & 90 & -168 & nef \\ 
44 & ${\bf WSP}(6,9,2,11,2,6,12|2k,24-2k)[24]$ & 6 & 114 & -216 & nef \\ 
45 & ${\bf WSP}(4,10,2,11,3,6,12|2k,24-2k)[24]$ & 7 & 55 & -96 & nonRG 
\\ 
46 & ${\bf WSP}(4,10,2,11,1,8,12|2k,24-2k)[24]$ & 3 & 99 & -192 \\ 
47 & ${\bf WSP}(2,11,2,11,4,6,12|2k,24-2k)[24]$ & 8 & 164 & -312 & nef 
\\ 
48 & ${\bf WSP}(2,11,2,11,2,8,12|2k,24-2k)[24]$ & 3 & 243 & -480 & nef 
\\ 
49 & ${\bf WSP}(4,12,2,13,4,7,14|2k,28-2k)[28]$ & 8 & 80 & -144 \\ 
50 & ${\bf WSP}(3,6,6,10,10,10,15|2k,30-2k)[30]$ & 21 & 21 & 0 & 
nef, $ T^2 \times K3 $ \\ 
51 & ${\bf WSP}(2,3,10,10,10,10,15|10,20)[30]$ & 17 & 41 & -48 \\ 
52 & ${\bf WSP}(6,12,5,6,6,10,15|6k,30-6k)[30]$ & 21 & 21 & 0 & nef,
$ T^2 \times K3 $ \\ 
53 & ${\bf WSP}(6,12,2,5,10,10,15|2k,30-2k)[30]$ & 17 & 41 & -48 \\ 
54 & ${\bf WSP}(6,12,1,6,10,10,15|2k,30-2k)[30]$ & 7 & 55 & -96 & 
nonRG \\ 
55 & ${\bf WSP}(2,14,3,6,10,10,15|2k,30-2k)[30]$ & 7 & 55 & -96 \\ 
56 & ${\bf WSP}(6,12,2,14,5,6,15|2k,30-2k)[30]$ & 7 & 55 & -96 & nonRG 
\\ 
57 & ${\bf WSP}(6,12,2,14,1,10,15|2k,30-2k)[30]$ & 5 & 101 & -192 \\ 
58 & ${\bf WSP}(2,14,2,14,3,10,15|2k,30-2k)[30]$ & 5 & 101 & -192 \\ 
59 & ${\bf WSP}(12,8,4,9,9,12,18|6k,36-6k)[36]$ & 21 & 21 & 0 & 
nef, $ T^2 \times K3 $ \\ 
60 & ${\bf WSP}(12,8,1,9,12,12,18|6k,36-6k)[36]$ & 10 & 46 & -72 \\ 
61 & ${\bf WSP}(12,8,6,15,4,9,18|6k,36-6k)[36]$ & 8 & 44 & -72 & nonRG 
\\ 
62 & ${\bf WSP}(12,8,6,15,1,12,18|6k,36-6k)[36]$ & 13 & 49 & -72 & 
nonRG \\ 
63 & ${\bf WSP}(4,16,4,9,9,12,18|2k,36-2k)[36]$ & 21 & 21 & 0 & nef,
$ T^2 \times K3 $ \\ 
64 & ${\bf WSP}(4,16,1,9,12,12,18|2k,36-2k)[36]$ & 20 & 56 & -72 \\ 
65 & ${\bf WSP}(6,15,4,16,4,9,18|2k,36-2k)[36]$ & 14 & 50 & -72 \\ 
66 & ${\bf WSP}(6,15,4,16,1,12,18|2k,36-2k)[36]$ & 5 & 77 & -144 \\ 
67 & ${\bf WSP}(12,8,2,17,6,9,18|2k,36-2k)[36]$ & 11 & 53 & -84 \\ 
68 & ${\bf WSP}(12,8,2,17,3,12,18|2k,36-2k)[36]$ & 13 & 49 & -72 & 
nonRG \\ 
69 & ${\bf WSP}(2,17,2,9,12,12,18|2k,36-2k)[36]$ & 16 & 100 & -168 \\ 
70 & ${\bf WSP}(6,15,2,17,2,12,18|2k,36-2k)[36]$ & 5 & 185 & -360 & 
nef \\ 
71 & ${\bf WSP}(4,16,2,17,6,9,18|2k,36-2k)[36]$ & 13 & 73 & -120 & 
nonRG \\ 
72 & ${\bf WSP}(4,16,2,17,3,12,18|2k,36-2k)[36]$ & 5 & 77 & -144 & 
nonRG \\ 
73 & ${\bf WSP}(2,17,2,17,4,12,18|2k,36-2k)[36]$ & 7 & 271 & -528 & 
nef \\ 
74 & ${\bf WSP}(10,15,10,15,2,8,20|10k,40-10k)[40]$ & 7 & 63 & -112 & 
nef \\ 
75 & ${\bf WSP}(10,15,8,16,1,10,20|2k,40-2k)[40]$ & 13 & 49 & -72 \\ 
76 & ${\bf WSP}(10,15,4,18,5,8,20|2k,40-2k)[40]$ & 19 & 27 & -16 \\ 
77 & ${\bf WSP}(10,15,2,19,4,10,20|2k,40-2k)[40]$ & 12 & 96 & -168 & 
nef \\ 
78 & ${\bf WSP}(8,16,2,19,5,10,20|2k,40-2k)[40]$ & 13 & 49 & -72 \\ 
79 & ${\bf WSP}(2,19,2,19,8,10,20|2k,40-2k)[40]$ & 11 & 227 & -432 & 
nef \\ 
80 & ${\bf WSP}(1,6,14,14,14,14,21|14,28)[42]$ & 23 & 47 & -48 \\ 
81 & ${\bf WSP}(6,18,6,18,1,14,21|6k,42-6k)[42]$ & 15 & 63 & -96 \\ 
82 & ${\bf WSP}(2,20,6,7,14,14,21|2k,42-2k)[42]$ & 23 & 47 & -48 \\ 
83 & ${\bf WSP}(6,18,2,20,3,14,21|2k,42-2k)[42]$ & 15 & 63 & -96 \\ 
84 & ${\bf WSP}(6,21,1,12,16,16,24|2k,48-2k)[48]$ & 20 & 56 & -72 \\ 
85 & ${\bf WSP}(8,20,6,21,1,16,24|2k,48-2k)[48]$ & 8 & 68 & -120 \\ 
86 & ${\bf WSP}(6,21,6,21,2,16,24|6k,48-6k)[48]$ & 7 & 127 & -240 & 
nef \\ 
87 & ${\bf WSP}(6,21,4,22,3,16,24|2k,48-2k)[48]$ & 17 & 41 & -48 \\ 
88 & ${\bf WSP}(2,23,3,12,16,16,24|2k,48-2k)[48]$ & 20 & 56 & -72 \\ 
89 & ${\bf WSP}(12,18,2,23,1,16,24|2k,48-2k)[48]$ & 9 & 129 & -240 \\ 
90 & ${\bf WSP}(8,20,2,23,3,16,24|2k,48-2k)[48]$ & 8 & 68 & -120 & 
nonRG \\ 
91 & ${\bf WSP}(6,21,2,23,8,12,24|2k,48-2k)[48]$ & 16 & 112 & -192 \\ 
92 & ${\bf WSP}(6,21,2,23,4,16,24|2k,48-2k)[48]$ & 8 & 164 & -312 & 
nonRG \\ 
93 & ${\bf WSP}(2,23,2,23,6,16,24|2k,48-2k)[48]$ & 9 & 321 & -624 & 
nef \\ 
94 & ${\bf WSP}(3,12,15,20,20,20,30|10k,60-10k)[60]$ & 21 & 21 & 0 & 
nef, $ T^2 \times K3 $ \\ 
95 & ${\bf WSP}(12,24,4,15,15,20,30|6k,60-6k)[60]$ & 21 & 21 & 0 & nef,
$ T^2 \times K3 $ \\ 
96 & ${\bf WSP}(10,25,3,12,20,20,30|10k,60-10k)[60]$ & 23 & 23 & 0 \\ 
97 & ${\bf WSP}(12,24,10,25,4,15,30|2k,60-2k)[60]$ & 29 & 29 & 0 \\ 
98 & ${\bf WSP}(6,27,10,12,15,20,30|6k,60-6k)[60]$ & 15 & 39 & -48 \\ 
99 & ${\bf WSP}(6,27,5,12,20,20,30|2k,60-2k)[60]$ & 23 & 23 & 0 \\ 
100 & ${\bf WSP}(6,27,2,15,20,20,30|2k,60-2k)[60]$ & 31 & 55 & -48 \\ 
101 & ${\bf WSP}(12,24,6,27,1,20,30|6k,60-6k)[60]$ & 17 & 65 & -96 \\ 
102 & ${\bf WSP}(10,25,6,27,10,12,30|2k,60-2k)[60]$ & 13 & 49 & -72 \\ 
103  & ${\bf WSP}(10,25,6,27,2,20,30|2k,60-2k)[60]$ & 10 & 106 & -192 
& nonRG \\ 
104 & ${\bf WSP}(6,27,6,27,4,20,30|6k,60-6k)[60]$ & 11 & 107 & -192 & 
nef \\ 
105 & ${\bf WSP}(4,28,3,15,20,20,30|2k,60-2k)[60]$ & 31 & 31 & 0 \\ 
106 & ${\bf WSP}(10,25,4,28,3,20,30|2k,60-2k)[60]$ & 10 & 46 & -72 & 
nonRG \\ 
107 & ${\bf WSP}(6,27,4,28,10,15,30|2k,60-2k)[60]$ & 25 & 37 & -24 & 
nonRG \\ 
108 & ${\bf WSP}(6,27,4,28,5,20,30|2k,60-2k)[60]$ & 10 & 46 & -72 & 
nonRG \\ 
109 & ${\bf WSP}(2,29,12,12,15,20,30|2k,60-2k)[60]$ & 23 & 47 & -48 \\ 
110 & ${\bf WSP}(2,29,4,15,20,20,30|2k,60-2k)[60]$ & 26 & 86 & -120 \\ 
111 & ${\bf WSP}(12,24,2,29,3,20,30|2k,60-2k)[60]$ & 17 & 65 & -96 \\ 
112 & ${\bf WSP}(10,25,2,29,12,12,30|2k,60-2k)[60]$ & 25 & 85 & -120 \\ 
113 & ${\bf WSP}(10,25,2,29,4,20,30|2k,60-2k)[60]$ & 11 & 155 & -188 & 
nef \\ 
114 & ${\bf WSP}(6,27,2,29,6,20,30|2k,60-2k)[60]$ & 10 & 178 & -336 \\ 
115 & ${\bf WSP}(4,28,2,29,12,15,30|2k,60-2k)[60]$ & 17 & 101 & -168 \\ 
116 & ${\bf WSP}(18,27,4,34,1,24,36|2k,72-2k)[72]$ & 14 & 98 & -168 \\ 
117 & ${\bf WSP}(18,27,2,35,8,18,36|2k,72-2k)[72]$ & 19 & 91 & -144 & 
nef \\ 
118 & ${\bf WSP}(18,27,2,35,2,24,36|2k,72-2k)[72]$ & 14 & 242 & -456 \\ 
119 & ${\bf WSP}(6,33,2,35,8,24,36|2k,72-2k)[72]$ & 15 & 183 & -336 \\ 
120 & ${\bf WSP}(4,34,2,35,9,24,36|2k,72-2k)[72]$ & 14 & 98 & -168 & 
nonRG \\ 
121 & ${\bf WSP}(12,36,1,21,28,28,42|2k,84-2k)[84]$ & 45 & 45 & 0 \\ 
122 & ${\bf WSP}(14,35,12,36,1,28,42|2k,84-2k)[84]$ & 15 & 63 & -96 \\ 
123 & ${\bf WSP}(6,39,4,21,28,28,42|2k,84-2k)[84]$ & 41 & 41 & 0 \\ 
124 & ${\bf WSP}(14,35,6,39,4,28,42|2k,84-2k)[84]$ & 16 & 76 & -120 & 
nonRG \\ 
125 & ${\bf WSP}(4,40,12,21,21,28,42|2k,84-2k)[84]$ & 21 & 21 & 0 & nef,
$ T^2 \times K3 $ \\ 
126 & ${\bf WSP}(14,35,4,40,12,21,42|2k,84-2k)[84]$ & 35 & 35 & 0 \\ 
127 & ${\bf WSP}(2,41,6,21,28,28,42|2k,84-2k)[84]$ & 40 & 76 & -72 \\ 
128 & ${\bf WSP}(14,35,2,41,6,28,42|2k,84-2k)[84]$ & 15 & 147 & -264 & 
nef \\ 
129 & ${\bf WSP}(12,36,2,41,14,21,42|2k,84-2k)[84]$ & 34 & 58 & -48 \\ 
130 & ${\bf WSP}(12,36,2,41,7,28,42|2k,84-2k)[84]$ & 15 & 63 & -96 \\ 
131 & ${\bf WSP}(2,41,2,41,12,28,42|2k,84-2k)[84]$ & 11 & 491 & -960 & 
nef \\ 
132 & ${\bf WSP}(30,20,18,36,1,30,45|6k,90-6k)[90]$ & 29 & 41 & -24 & 
nonRG \\ 
133 & ${\bf WSP}(18,36,10,40,1,30,45|2k,90-2k)[90]$ & 17 & 65 & -96 \\ 
134 & ${\bf WSP}(30,20,2,44,9,30,45|2k,90-2k)[90]$ & 29 & 41 & -24 & 
nonRG \\ 
135 & ${\bf WSP}(18,36,2,44,5,30,45|2k,90-2k)[90]$ & 17 & 65 & -96 \\ 
136 & ${\bf WSP}(10,40,2,44,9,30,45|2k,90-2k)[90]$ & 17 & 65 & -96 \\ 
137 & ${\bf WSP}(24,36,6,45,1,32,48|6k,96-6k)[96]$ & 24 & 84 & -120 \\ 
138 & ${\bf WSP}(24,36,2,47,3,32,48|2k,96-2k)[96]$ & 24 & 84 & -120 \\ 
139 & ${\bf WSP}(6,45,2,47,12,32,48|2k,96-2k)[96]$ & 18 & 222 & -408 \\ 
140 & ${\bf WSP}(30,45,1,24,40,40,60|10k,120-10k)[120]$ & 39 & 39 & 0 
& nonRG \\ 
141 & ${\bf WSP}(24,48,10,55,3,40,60|2k,120-2k)[120]$ & 29 & 29 & 0 \\ 
142 & ${\bf WSP}(30,45,8,56,1,40,60|2k,120-2k)[120]$ & 24 & 84 & -120 \\ 
143 & ${\bf WSP}(30,45,6,57,2,40,60|6k,120-6k)[120]$ & 23 & 143 & -240 
\\ 
144 & ${\bf WSP}(24,48,6,57,5,40,60|6k,120-6k)[120]$ & 29 & 29 & 0 \\ 
145 & ${\bf WSP}(10,55,6,57,12,40,60|2k,120-2k)[120]$ & 22 & 82 & -120 
\\ 
146 & ${\bf WSP}(30,45,4,58,3,40,60|2k,120-2k)[120]$ & 29 & 53 & -48 \\ 
147 & ${\bf WSP}(6,57,4,58,15,40,60|2k,120-2k)[120]$ & 29 & 53 & -48 & 
nonRG \\ 
148 & ${\bf WSP}(2,59,15,24,40,40,60|2k,120-2k)[120]$ & 39 & 39 & 0 \\ 
149 & ${\bf WSP}(30,45,2,59,20,24,60|2k,120-2k)[120]$ & 33 & 69 & -72 \\ 
150 & ${\bf WSP}(30,45,2,59,4,40,60|2k,120-2k)[120]$ & 24 & 204 & -360 
\\ 
151 & ${\bf WSP}(10,55,2,59,24,30,60|2k,120-2k)[120]$ & 33 & 141 & 
-216 \\ 
152 & ${\bf WSP}(8,56,2,59,15,40,60|2k,120-2k)[120]$ & 24 & 84 & -120 \\ 
153 & ${\bf WSP}(10,65,4,68,28,35,70|2k,140-2k)[140]$ & 47 & 47 & 0 \\ 
154 & ${\bf WSP}(28,56,2,69,20,35,70|2k,140-2k)[140]$ & 53 & 53 & 0 \\ 
155 & ${\bf WSP}(2,77,12,39,52,52,78|2k,156-2k)[156]$ & 71 & 71 & 0 \\ 
156 & ${\bf WSP}(26,65,2,77,12,52,78|2k,156-2k)[156]$ & 23 & 143 & 
-240 & nef \\ 
157 & ${\bf WSP}(42,63,12,78,1,56,84|6k,168-6k)[168]$ & 38 & 74 & -72 \\ 
158 & ${\bf WSP}(42,63,8,80,3,56,84|2k,168-2k)[168]$ & 39 & 39 & 0 \\ 
159 & ${\bf WSP}(42,63,6,81,4,56,84|6k,168-6k)[168]$ & 33 & 105 & -144 
\\ 
160 & ${\bf WSP}(8,80,6,81,21,56,84|2k,168-2k)[168]$ & 39 & 39 & 0 \\ 
161 & ${\bf WSP}(42,63,2,83,6,56,84|2k,168-2k)[168]$ & 34 & 190 & -312 
\\ 
162 & ${\bf WSP}(12,78,2,83,21,56,84|2k,168-2k)[168]$ & 38 & 74 & -72 \\ 
163 & ${\bf WSP}(6,81,2,83,24,56,84|2k,168-2k)[168]$ & 23 & 335 & -624 
\\ 
164 & ${\bf WSP}(18,81,2,89,20,60,90|2k,180-2k)[180]$ & 42 & 150 & 
-216 \\ 
165 & ${\bf WSP}(54,81,2,107,8,72,108|2k,216-2k)[216]$ & 48 & 180 & 
-264 \\ 
166 & ${\bf WSP}(20,100,2,109,44,55,110|2k,220-2k)[220]$ & 71 & 71 & 0 
\\ 
167 & ${\bf WSP}(48,96,30,105,1,80,120|6k,240-6k)[240]$ & 53 & 53 & 0 \\ 
168 & ${\bf WSP}(48,96,2,119,15,80,120|2k,240-2k)[240]$ & 53 & 53 & 0 \\ 
169 & ${\bf WSP}(30,105,2,119,24,80,120|2k,240-2k)[240]$ & 50 & 134 & 
-168 \\ 
170 & ${\bf WSP}(66,99,6,129,8,88,132|6k,264-6k)[264]$ & 57 & 81 & -48 
\\ 
171 & ${\bf WSP}(78,117,24,144,1,104,156|6k,312-6k)[312]$ & 69 & 69 & 
0 \\ 
172 & ${\bf WSP}(78,117,2,155,12,104,156|2k,312-2k)[312]$ & 66 & 174 & 
-216 \\ 
173 & ${\bf WSP}(24,144,2,155,39,104,156|2k,312-2k)[312]$ & 69 & 69 & 
0 \\ 
174 & ${\bf WSP}(14,161,2,167,48,112,168|2k,336-2k)[336]$ & 47 & 287 & 
-480 \\ 
175 & ${\bf WSP}(14,203,6,207,60,140,210|2k,420-2k)[420]$ & 59 & 131 & 
-144 \\ 
176 & ${\bf WSP}(150,225,2,299,24,200,300|2k,600-2k)[600]$ & 119 & 167 & 
-96 \\ 
177 & ${\bf WSP}(42,441,2,461,132,308,462|2k,924-2k)[924]$ & 137 & 257 & 
-240 \\ 
178 & ${\bf WSP}(1,2,1,2,1,2,1|k,5-k)[5]$ & 1 & 85 & -168 \\ 
179 & ${\bf WSP}(1,1,2,2,2,2,2|2,4)[6]$ & 1 & 73 & -144 \\ 
180 & ${\bf WSP}(2,3,2,3,2,2,2|2k,8-2k)[8]$ & 2 & 86 & -168 & nef \\ 
181 & ${\bf WSP}(2,3,2,3,2,3,1|k,8-k)[8]$ & 2 & 58 & -112 \\ 
182 & ${\bf WSP}(3,2,3,2,3,2,3|k,9-k)[9]$ & 8 & 35 & -54 & nef \\ 
183 & ${\bf WSP}(3,2,1,3,3,3,3|3,6)[9]$ & 2 & 62 & -120 & nonRG \\ 
184 & ${\bf WSP}(3,2,3,2,1,4,3|k,9-k)[9]$ & 2 & 56 & -108 \\ 
185 & ${\bf WSP}(1,4,1,3,3,3,3|k,9-k)[9]$ & 8 & 68 & -120 \\ 
186 & ${\bf WSP}(3,2,1,4,1,4,3|k,9-k)[9]$ & 4 & 76 & -144 \\ 
187 & ${\bf WSP}(1,4,1,4,1,4,3|k,9-k)[9]$ & 2 & 110 & -216 \\ 
188 & ${\bf WSP}(3,3,3,3,4,4,4|k,12-k)[12]$ & 21 & 21 & 0 & nef,
$ T^2 \times K3 $ \\ 
189 & ${\bf WSP}(2,3,3,4,4,4,4|2k,12-2k)[12]$ & 11 & 35 & -48 \\ 
190 & ${\bf WSP}(1,3,4,4,4,4,4|4,8)[12]$ & 2 & 62 & -120 \\ 
191 & ${\bf WSP}(2,5,3,3,3,4,4|k,12-k)[12]$ & 4 & 40 & -72 \\
192 & ${\bf WSP}(2,5,2,3,4,4,4|2k,12-2k)[12]$ & 2 & 62 & -120 & nonRG \\ 
193 & ${\bf WSP}(2,5,1,4,4,4,4|2k,12-2k)[12]$ & 9 & 57 & -96 \\ 
194 & ${\bf WSP}(2,5,2,5,3,3,4|k,12-k)[12]$ & 5 & 41 & -72 \\ 
195 & ${\bf WSP}(2,5,2,5,2,4,4|2k,12-2k)[12]$ & 5 & 101 & -192 & nef \\ 
196 & ${\bf WSP}(2,5,2,5,2,5,3|k,12-k)[12]$ & 3 & 57 & -108 \\ 
197 & ${\bf WSP}(3,6,3,3,5,5,5|k,15-k)[15]$ & 21 & 21 & 0 & nef,
$ T^2 \times K3 $ \\ 
198 & ${\bf WSP}(3,6,1,5,5,5,5|k,15-k)[15]$ & 17 & 41 & -48 \\ 
199 & ${\bf WSP}(3,6,1,7,3,5,5|k,15-k)[15]$ & 7 & 55 & -96 & nonRG \\ 
200 & ${\bf WSP}(3,6,1,7,1,7,5|k,15-k)[15]$ & 5 & 101 & -192 \\ 
201 & ${\bf WSP}(4,6,2,7,2,7,4|2k,16-2k)[16]$ & 8 & 104 & -192 & nef \\ 
202 & ${\bf WSP}(4,8,4,8,2,9,5|2k,20-2k)[20]$ & 13 & 49 & -72 \\ 
203 & ${\bf WSP}(1,10,3,7,7,7,7|k,21-k)[21]$ & 23 & 47 & -48 \\
204 & ${\bf WSP}(3,9,3,9,1,10,7|k,21-k)[21]$ & 15 & 63 & -96 \\ 
205 & ${\bf WSP}(6,9,3,6,8,8,8|2k,24-2k)[24]$ & 21 & 21 & 0 & nef,
$ T^2 \times K3 $ \\ 
206 & ${\bf WSP}(6,9,1,8,8,8,8|2k,24-2k)[24]$ & 24 & 36 & -24 \\ 
207 & ${\bf WSP}(6,9,6,9,4,6,8|6k,24-6k)[24]$ & 7 & 55 & -96 & nef \\ 
208 & ${\bf WSP}(6,9,6,9,2,8,8|2k,24-2k)[24]$ & 19 & 43 & -48 & nef \\ 
209 & ${\bf WSP}(6,9,4,10,3,8,8|2k,24-2k)[24]$ & 9 & 33 & -48 \\ 
210 & ${\bf WSP}(6,9,6,9,4,10,4|2k,24-2k)[24]$ & 20 & 32 & -24 & nef \\ 
211 & ${\bf WSP}(2,11,3,8,8,8,8|2k,24-2k)[24]$ & 24 & 36 & -24 \\ 
212 & ${\bf WSP}(6,9,2,11,6,6,8|2k,24-2k)[24]$ & 8 & 68 & -120 & nonRG 
\\ 
213 & ${\bf WSP}(6,9,2,11,4,8,8|2k,24-2k)[24]$ & 10 & 70 & -120 & 
nonRG \\ 
214 & ${\bf WSP}(6,9,4,10,2,11,6|2k,24-2k)[24]$ & 10 & 70 & -120 & 
nonRG \\ 
215 & ${\bf WSP}(2,11,2,11,6,8,8|2k,24-2k)[24]$ & 11 & 131 & -240 & 
nef \\ 
216 & ${\bf WSP}(4,10,2,11,2,11,8|2k,24-2k)[24]$ & 9 & 153 & -288 & 
nef \\ 
217 & ${\bf WSP}(12,8,2,17,9,12,12|2k,36-2k)[36]$ & 10 & 46 & -72 \\ 
218 & ${\bf WSP}(12,8,6,15,2,17,12|2k,36-2k)[36]$ & 17 & 77 & -120 & 
nonRG \\ 
219 & ${\bf WSP}(4,16,2,17,9,12,12|2k,36-2k)[36]$ & 20 & 56 & -72 \\ 
220 & ${\bf WSP}(6,15,4,16,2,17,12|2k,36-2k)[36]$ & 13 & 109 & -192 & 
nef \\ 
221 & ${\bf WSP}(10,15,10,15,4,18,8|2k,40-2k)[40]$ & 31 & 23 & 16 & 
nef \\ 
222 & ${\bf WSP}(10,15,8,16,2,19,10|2k,40-2k)[40]$ & 23 & 59 & -72 \\ 
223 & ${\bf WSP}(15,10,9,18,1,22,15|k,45-k)[45]$ & 29 & 41 & -24 & 
nonRG \\ 
224 & ${\bf WSP}(9,18,5,20,1,22,15|k,45-k)[45]$ & 17 & 65 & -96 \\ 
225 & ${\bf WSP}(6,21,6,21,4,22,16|2k,48-2k)[48]$ & 31 & 55 & -48 & 
nef \\ 
226 & ${\bf WSP}(6,21,2,23,12,16,16|2k,48-2k)[48]$ & 26 & 86 & -120 \\ 
227 & ${\bf WSP}(8,20,6,21,2,23,16|2k,48-2k)[48]$ & 18 & 102 & -168 \\ 
228 & ${\bf WSP}(12,18,2,23,2,23,16|2k,48-2k)[48]$ & 13 & 229 & -432 & 
nef  \\ 
229 & ${\bf WSP}(6,27,12,15,20,20,20|2k,60-2k)[60]$ & 21 & 21 & 0 & nef,
$ T^2 \times K3 $ \\ 
230 & ${\bf WSP}(10,25,6,27,12,20,20|2k,60-2k)[60]$ & 31 & 31 & 0 \\ 
231 & ${\bf WSP}(6,27,4,28,15,20,20|2k,60-2k)[60]$ & 31 & 31 & 0 \\ 
232 & ${\bf WSP}(10,25,6,27,4,28,20|2k,60-2k)[60]$ & 22 & 58 & -72 & 
nonRG \\ 
233 & ${\bf WSP}(12,24,6,27,2,29,20|2k,60-2k)[60]$ & 21 & 117 & -192 \\ 
234 & ${\bf WSP}(18,27,4,34,2,35,24|2k,72-2k)[72]$ & 22 & 130 & -216 \\ 
235 & ${\bf WSP}(12,36,2,41,21,28,28|2k,84-2k)[84]$ & 45 & 45 & 0 \\ 
236 & ${\bf WSP}(14,35,12,36,2,41,28|2k,84-2k)[84]$ & 37 & 85 & -96 \\ 
237 & ${\bf WSP}(24,36,6,45,2,47,32|2k,96-2k)[96]$ & 26 & 158 & -264 \\ 
238 & ${\bf WSP}(24,48,10,55,6,57,40|2k,120-2k)[120]$ & 49 & 49 & 0 \\ 
239 & ${\bf WSP}(30,45,6,57,4,58,40|2k,120-2k)[120]$ & 43 & 67 & -48 \\ 
240 & ${\bf WSP}(30,45,2,59,24,40,40|2k,120-2k)[120]$ & 55 & 55 & 0 \\ 
241 & ${\bf WSP}(30,45,8,56,2,59,40|2k,120-2k)[120]$ & 34 & 118 & -168 
\\ 
242 & ${\bf WSP}(42,63,8,80,6,81,56|2k,168-2k)[168]$ & 55 & 55 & 0 \\ 
243 & ${\bf WSP}(42,63,12,78,2,83,56|2k,168-2k)[168]$ & 46 & 106 & 
-120 & nonRG \\ 
244 & ${\bf WSP}(48,96,30,105,2,119,80|2k,240-2k)[240]$ & 89 & 89 & 0 \\ 
245 & ${\bf WSP}(78,117,24,144,2,155,104|2k,312-2k)[312]$ & 97 & 97 & 
0 \\ 
246 & ${\bf WSP}(1,2,2,2,2,2,2,2,3|2,4,2,4)[6]$ & 0 & 84 & -168 \\ 
247 & ${\bf WSP}(2,3,2,3,2,3,2,3,4|k,8-k,2,6)[8]$ & 1 & 53 & -104 \\ 
    & ${\bf WSP}(2,3,2,3,2,3,2,3,4|3,5,4,4)[8]$ & 1 & 53 & -104 \\ 
248 & ${\bf WSP}(3,3,4,4,4,4,4,4,6|2k,12-2k,4,8)[12]$ & 21 & 21 & 0 & 
$ T^2 \times K3 $ \\ 
249 & ${\bf WSP}(2,5,3,4,4,4,4,4,6|2k,12-2k,4,8)[12]$ & 2 & 62 & -120 & 
nonRG \\ 
250 & ${\bf WSP}(2,5,2,5,4,4,4,4,6|2k,12-2k,2l,12-2l)[12]$ & 7 & 79 & 
-144 \\ 
251 & ${\bf WSP}(6,9,6,9,6,8,8,8,12|2k,24-2k,6l,24-6l)[24]$ & 21 & 21 
& 0 & nef, $ T^2 \times K3 $ \\ 
252 & ${\bf WSP}(6,9,6,9,4,10,8,8,12|2k,24-2k,6l,24-6l)[24]$ & 11 & 35 & 
-48 \\ 
253 & ${\bf WSP}(6,9,2,11,8,8,8,8,12|2k,24-2k,4l,24-4l)[24]$ & 16 & 52 & 
-72 \\ 
254 & ${\bf WSP}(1,1,1,1,1,1,1,1,1|1,2,1,2)[3]$ & 0 & 84 & -168 \\
\label{ResultsTable}
\end{longtable}
\normalsize

\section*{Acknowledgements}

R.G. and M.K. acknowledge financial support from the Austrian Research 
Funds FWF under grant number P18679-N16.


\end{document}